\documentclass[useAMS,usenatbib]{mnras}

\usepackage{newtxtext,newtxmath}
\usepackage[T1]{fontenc}
\usepackage{ae,aecompl}
\usepackage{ulem}
\usepackage{color}
\usepackage{multirow}
\usepackage{graphicx}	
\usepackage{amsmath}	
\usepackage{amssymb}	
\usepackage{xfrac}
\usepackage{float}
\usepackage{caption}
\usepackage{subcaption}
\usepackage{natbib}
\setkeys{Gin}{draft=false}
\title[Thermohaline mixing in globular cluster red giants]{A phenomenological modification of thermohaline mixing in globular cluster red giants}

\author[Henkel et al.]{
Kate Henkel,$^{1}$\thanks{E-mail: kate.henkel@monash.edu}
Amanda I. Karakas,$^{1}$
John C. Lattanzio$^{1}$
\\
$^{1}$Monash Centre for Astrophysics, School of Physics \& Astronomy, Monash University, Clayton 3800, Victoria, Australia\\
}

\date{Accepted XXX. Received YYY; in original form ZZZ}

\pubyear{2017}

\begin{document}
\label{firstpage}
\pagerange{\pageref{firstpage}--\pageref{lastpage}}
\maketitle

\begin{abstract}
Thermohaline mixing is a favoured mechanism for the so-called ``extra mixing'' on the red giant branch of low-mass stars. The mixing is triggered by the molecular weight inversion created above the hydrogen shell during first dredge-up when helium-3 burns via $^3$He($^3$He,2p)$^4$He. The standard 1D diffusive mixing scheme cannot simultaneously match carbon and lithium abundances to NGC 6397 red giants. We investigate two modifications to the standard scheme: (1) an advective two stream mixing algorithm, and (2) modifications to the standard 1D thermohaline mixing formalism. We cannot simultaneously match carbon and lithium abundances using our two stream mixing approach. However we develop a modified diffusive scheme with an explicit temperature dependence that can simultaneously fit carbon and lithium abundances to NGC 6397 stars. Our modified diffusive scheme induces mixing that is faster than the standard theory predicts in the hotter part of the thermohaline region and mixing that is slower in the cooler part. Our results infer that the extra mixing mechanism needs further investigation and more observations are required, particularly for stars in different clusters spanning a range in metallicity.
\end{abstract}




\section{Introduction}
\label{sec:intro}

Mixing in stars is a complicated subject and modelling mixing events proves to be a difficult task. It is perhaps not surprising that there is observational evidence for mixing due to processes not predicted by theory. This is particularly the case for post-main sequence stars ascending the Red Giant Branch (RGB).

As a low-mass ($\lesssim 2.5 \text{M}_{\odot}$) star ascends the RGB, the energy generated by the hydrogen burning shell (H shell) is transported to the surface easily via the convective envelope. This drives further expansion and deepens the mass location of the base of the convective envelope. The envelope eventually makes contact with the ashes of main sequence hydrogen burning and mixes these ashes to the surface. This process is called first dredge-up (FDU) and is the first major mixing event in a stars life \citep{iben64}.

FDU leaves behind a hydrogen abundance discontinuity that becomes very important for subsequent evolution \citep{iben64}. This discontinuity is believed to inhibit any extra mixing between the H shell and envelope \citep{mestel53,kippwei90,chaname05}. As the H shell converts hydrogen into helium, the inert helium core grows in mass. During core growth the H shell will move out (in mass) until it eventually reaches the hydrogen discontinuity left behind by FDU. The amount of available fuel and opacity of the H shell increases and the rate of hydrogen burning drops, which causes a drop in luminosity but a rise in effective temperature. The star appears to retrace its path in the Hertzsprung-Russell diagram (HRD) during this structural change. After the H shell passes through the hydrogen discontinuity, the luminosity will grow with core mass and the star will move up the RGB again almost along its original path. More stars are observed in this region of the HRD, producing a bump in the luminosity function (LFB, also known as the ``RGB bump'' or simply ``bump''), which is a well-observed property of globular clusters \citep{kippwei90,riello03}. After the LFB, the hydrogen discontinuity is erased and it is thought that extra mixing can then occur \citep{lattanzio03,charbonnel07a,karakas14dawes}. Low-mass stars have sufficiently long RGB lifetimes to allow the H shell to connect to the convective envelope via extra mixing mechanisms. This connection allows abundance changes occurring below the convection zone to be observed on the surface.

Globular clusters are an ideal laboratory for studies of extra mixing in stars. This is because globular clusters contain hundreds of thousands of low-mass stars, and the globular cluster population covers a huge range in metallicity. Globular clusters are also known to harbour multiple populations \citep{gratton04,gratton12}, where these populations are correlated with spreads in the abundances of light elements \citep[e.g., O, Na,][]{cottrell81}. Carbon and lithium abundances are useful indicators of mixing episodes during stellar evolution because they burn at different temperatures, meaning they are independently sensitive to temperature changes in the star and therefore reflect the temperature and mixing history of stars. The abundances of carbon and lithium in particular trace the evolutionary history of a star within a globular cluster \citep{briley90,gilroy91,gratton00,shetrone03,gsmith03,weiss04,martell08,lind09}. This is especially evident for NGC 6397, where the lithium abundance has been shown to vary with luminosity and a noticeable drop in the abundance is evident after FDU \citep[e.g.,][]{lind09}. Furthermore, there are determinations of the carbon abundances for NGC 6397 \citep{briley90}, which makes this cluster the ideal place for us to start our investigation.

A popular theory that offers a solution to extra mixing is thermohaline mixing. First introduced by \citet{ulrich72} and then advanced further by \citet{kippenhahn80} in a general stellar context, an environment suitable for thermohaline mixing occurs naturally in low- to intermediate-mass stars after FDU \citep{eggleton08}. During core hydrogen burning, $^3$He is efficiently destroyed in the core but is produced outside the core where lower temperatures halt destruction. After FDU, $^3$He is homogenised in the outer layers and it is in this way that the surface value of $^3$He in low-mass stars increases. Once the outer layers are lost to the interstellar medium during later stages of evolution, the amount of $^3$He released is significant. This gives rise to the $^3$He abundance problem discussed by many authors since the mid 1990's \citep[e.g.,][]{hata95,olive95,eggleton06}, where the amount of $^3$He observed in the interstellar medium matches what is predicted from Big Bang nucleosynthesis but is much lower than predicted from enrichment by low-mass stellar models. This is discussed in more detail by \citet{eggleton06}. 

Therefore if Big Bang nucleosynthesis is correct, then the theory of low-mass stellar evolution must be incomplete and some process (or processes) must destroy $^3$He in low-mass stars. Helium-3 can undergo fusion via $^3$He($^3$He,2p)$^4$He once the H shell has advanced to regions where temperatures are high enough \citep[e.g.,][]{denissenkov04}. This fusion process creates more particles than it consumes and produces a mean molecular weight ($\mu$) inversion \citep{eggleton06}. \citet{charbonnel07a} linked this phenomenon to thermohaline mixing. A cooler layer of higher $\mu$ material sits atop a hotter, lower $\mu$ layer undergoing $^3$He fusion. ``Fingers'' of material from the hotter $^3$He fusion layer penetrate into the layer above, mixing material towards the surface.

There are other explanations of extra mixing in the literature. Most 1D stellar models do not include rotation, though it is known that stars rotate and this fact has been a source of discussion surrounding the (currently) unexplained mixing seen above the LFB. In three papers, \citet{charbonnel10}, \citet{lagarde11}, and \citet{lagarde12b} constructed a grid of models that included both rotation and thermohaline mixing, and combined these effects by adding the diffusion coefficients of each. This method does not account for any possible interaction between the two instabilities. This is discussed by \citet{cantiello10} who state that the diffusion coefficient for rotationally-induced instabilities is much lower than the value for thermohaline mixing by several orders of magnitude. \citet{canuto99} show that horizontal forces caused by rotation decrease the magnitude and efficiency of thermohaline mixing, though \citet{medrano14} found the opposite: the horizontal instabilities created by rotation increase the efficiency of thermohaline mixing in their 3D simulations. It is clear that there is much that is not understood in this area. \citet{maeder13} examine the interactions of instabilities (including thermohaline mixing) in rotating stars, an analysis that had been missing from the literature. Rotation alone cannot account for the extra mixing seen on the RGB \citep{chaname05,palacios06}, yet a combination of instabilities (such as rotation, gravity waves, and thermohaline mixing) could prove to be a candidate \citep{denissenkov09,denissenkov12}. This is discussed further in \S\ref{sec:discussion}.

Additionally, the stellar environment has consequences for thermohaline mixing, which was discovered in an oceanographic context that is incomparable (beyond a general sense) to stellar conditions. \citet{cantiello10} show that the dimensionless Prandtl number, $\sigma$, the ratio of the kinematic viscosity to the thermal diffusivity, is very small in stars at 10$^{-6}$. However, the Prandtl number in water (where thermohaline mixing was first discovered) is around 7 \citep{cantiello10}. Low Prandtl numbers (i.e. when $\sigma$ $\rightarrow$ 0) result in the (stellar) environment becoming unstable. This is shown in the simulations performed by \citet{bascoul07} where fluid environments with low Prandtl numbers are dynamically unstable and turbulent. Whether thermohaline mixing can survive in such turbulent conditions is still under debate \citep{busse78,merryfield95,bascoul07,cantiello10}.

One-dimensional models of thermohaline mixing also find discrepancies with observations. \citet{angelou15} attempted to match their stellar models to observations of lithium and carbon in globular cluster RGB stars. They could not find a simultaneous match, which led them to the conclusion that the diffusive mixing scheme adopted for thermohaline mixing may not be complete or the most appropriate. However, important benefits of the thermohaline mechanism as a possible solution to extra mixing are that it naturally begins at the LFB and the thermohaline environment naturally occurs during normal RGB evolution. Indeed, it is also possible that the thermohaline instability triggers other instabilities that may then dominate mixing (this is discussed further in \S\ref{sec:discussion}).

The aim of this study is to investigate models of transport by thermohaline mixing. In \S\ref{sec:methods} we describe our numerical methods. In \S\ref{sec:models} we discuss the stellar models used. In \S\ref{sec:results} we state our test cases and show the results we obtain, and in \S\ref{sec:discussion} we discuss the results and state our conclusions.



\section{Numerical Methods}
\label{sec:methods}

\subsection{The evolution code MONSTAR}
\label{subsec:evcode}

All stellar models have been run using MONSTAR, the Monash version of the Mt. Stromlo stellar evolution code, as described in \citet{angelou15}. High-temperature opacities are provided by the OPAL opacity tables \citep{iglesias96}. Also included are fixed-metal distribution (OPAL type-1) tables that have the solar mixture composition of \citet{grevesse93} and $\alpha$-element enhancement of Achim Weiss \citep{iglesias96}. Tables variable in C and O content (OPAL type-2) are based upon abundances of \citet{grevesse93}. MONSTAR uses low-temperature (below $10^4$ K) tables with variable C and N content from \citet{lederer09} \citep{karakas10b,angelou15}. The mass loss prescription on the RGB is from \citet{reimers75} with $\eta_R = 0.4$ \citep{constantino16}.

MONSTAR has a network of nine species ($^1$H, $^3$He, $^4$He, $^7$Be, $^7$Li, $^{12}$C, $^{13}$C, $^{14}$N, $^{16}$O, and a tenth, inert, species). All key reaction rates included in MONSTAR, as well as more detail on the reaction rates used, can be found in Table 1 of \citet{angelou15}. For the importance of timestepping and spatial resolution in MONSTAR (and evolution codes used by other groups) see \citet{lattanzio15}.

Convection is modelled according to the mixing length theory \citep[MLT;][]{bohm58} and we adopt $\alpha_{\rm MLT}=1.69$. Mixing of chemical species is modelled via a diffusion equation \citep{campbell08}. The Schwarzschild criterion is adopted to define convective boundaries. Also included is overshoot beyond the formal boundary. In \S\ref{subsubsec:overshoot} we describe our convective model in more detail.

\subsubsection{Convective overshoot}
\label{subsubsec:overshoot}

Convective overshoot is included following the procedure of \citet{herwig97}, which inserts an exponential decay in the velocity of overshooting material. Let $H_P$ be the pressure scale height, $f_{OS}$ a scaling factor, and the velocity scale-height $H_v$ be
\begin{equation}
H_v = f_{OS} H_P.
\label{eq:hv}
\end{equation}
We set $f_{OS}=0.14$ for material overshooting the convective boundary to match the observed location of FDU and the RGB bump for NGC 6397 RGB stars (discussed in more detail in \S\ref{subsubsec:fdubump}). The diffusion coefficient for overshooting material, $D_{OS}$, at a distance $z$ from the convective boundary is
\begin{equation}
D_{OS} = D_{conv} e^{-2z/H_v},
\label{eq:dos}
\end{equation}
where $D_{conv}$ is the diffusion coefficient at the convective boundary. Note that the only convective boundary in our models is at the base of the convective envelope.

The value of $f_{OS}$ in Equation~\ref{eq:hv} affects the depth of FDU and consequently the luminosity of the RGB bump. A high overshoot factor means that more material overshoots the convective envelope border and FDU is deeper. When the H shell advances in mass during normal RGB evolution, it will thus encounter the discontinuity caused by FDU sooner, resulting in a lower bump luminosity.

\subsubsection{Thermohaline mixing in MONSTAR}
\label{subsubsec:thmev}

The diffusion coefficient for thermohaline mixing is given by \citet{ulrich72} and \citet{kippenhahn80} as
\begin{equation}
\label{eq:Dthm}
D_t = C_t K (\phi / \delta) {{-\nabla_{\mu}} \over {(\nabla_{\rm ad} - \nabla)}}.
\end{equation}
The dimensionless parameter $C_t$ is formally given by
\begin{equation}
\label{eq:Cthm}
C_t = \frac{8}{3} \pi^2 \alpha^2,
\end{equation}
where $\alpha$ is the aspect ratio of thermohaline ``fingers'', $\phi = (\partial \ln \rho/\partial \ln \mu)_{P,T}$, $\delta=(\partial \ln \rho/\partial \ln T)_{P,\rho}$ ($\phi = \delta = 1$ for an ideal gas), $K=4acT^3/(3\kappa \rho^2 c_P)$ is the thermal diffusivity, $a$ is the radiation density constant, $\kappa$ is the Rosseland mean opacity, $c_P$ is the specific heat capacity, and all other symbols have their usual meanings. The value of $\alpha$ is not known a priori, hence we treat $C_t$ as a free parameter.

\subsubsection{Mixing timescales}
\label{subsubsec:mixtimescale}

In regions where mixing is modelled by a diffusion equation, a particle will travel with velocity $v$ (cm s$^{-1}$) over a mixing length $l$ (cm). The diffusion coefficient is related to $v$ and $l$ by
\begin{equation}
\label{eq:diff}
D \equiv \frac{1}{3} v l,
\end{equation}
where $D$ is in cm$^2$ s$^{-1}$. We calculate the timescale $\tau_{\rm thm}$ of mixing in this region by using the relation
\begin{equation}
\label{eq:tau}
\tau_{\rm thm} = \cfrac{l}{v},
\end{equation}
or
\begin{equation}
\label{eq:tauthm}
\tau_{\rm thm} \equiv {{l^2} \over {3 D_t}}.
\end{equation}

\subsubsection{Nuclear burning}
\label{subsubsec:nuclearburning}

Lithium-7, which burns at around $2 \times 10^6$ K, is a useful tracer of mixing and burning. This is especially the case for low-mass RGB stars that have extensive convective envelopes. The sudden drop in surface $^7$Li after FDU is one of the key observations that provides evidence for the existence of extra mixing \citep{charbonnel98,smiljanic09,lind09}. Beryllium-7, which captures an electron to form $^7$Li as part of the ppII chain, is an integral isotope in the Cameron-Fowler mechanism \citep[a means for producing $^7$Li in stellar interiors;][]{cameron71}. Beryllium-7 has a short half-life and is produced via alpha capture on $^3$He as the first reaction in the ppII and ppIII chains. Helium-3 itself is important because destruction of $^3$He via $^3$He($^3$He,2p)$^4$He is the reaction that produces a local decrease in $\mu$, thus driving thermohaline mixing.

Another key piece of evidence for extra mixing is observations of [C/Fe]. This ratio also shows a sharp decline after the completion of FDU, confirming that there is a non-canonical mixing event occurring in the stellar interior \citep{suntzeff81,suntzeff89,carbon82,trefzger83,langer86,briley90,gratton00,bellman01,shetrone03,martell08,shetrone10}.

\subsection{The post-processing nucleosynthesis code MONSOON}
\label{sec:nscode}

Our stellar models have been constructed using MONSTAR and then processed by the Monash post-processing nucleosynthesis code MONSOON \citep{cannon93,campbell08,karakas14dawes,karakas16}. MONSOON requires as input the radius, temperature, density, mixing length, and velocity as a function of interior mass and time. The reaction rates used in MONSOON are from the JINA REACLIB database \citep{cyburt10}.

Note that in each code we employ different mixing algorithms. In MONSTAR we employ a diffusive mixing scheme as outlined in \S\ref{subsubsec:thmev}. In MONSOON we employ an advective mixing scheme as outlined in \S\ref{subsubsec:twostream}.

\subsubsection{Two stream advective mixing}
\label{subsubsec:twostream}

\begin{figure}
   \centering
   \includegraphics[width=\columnwidth]{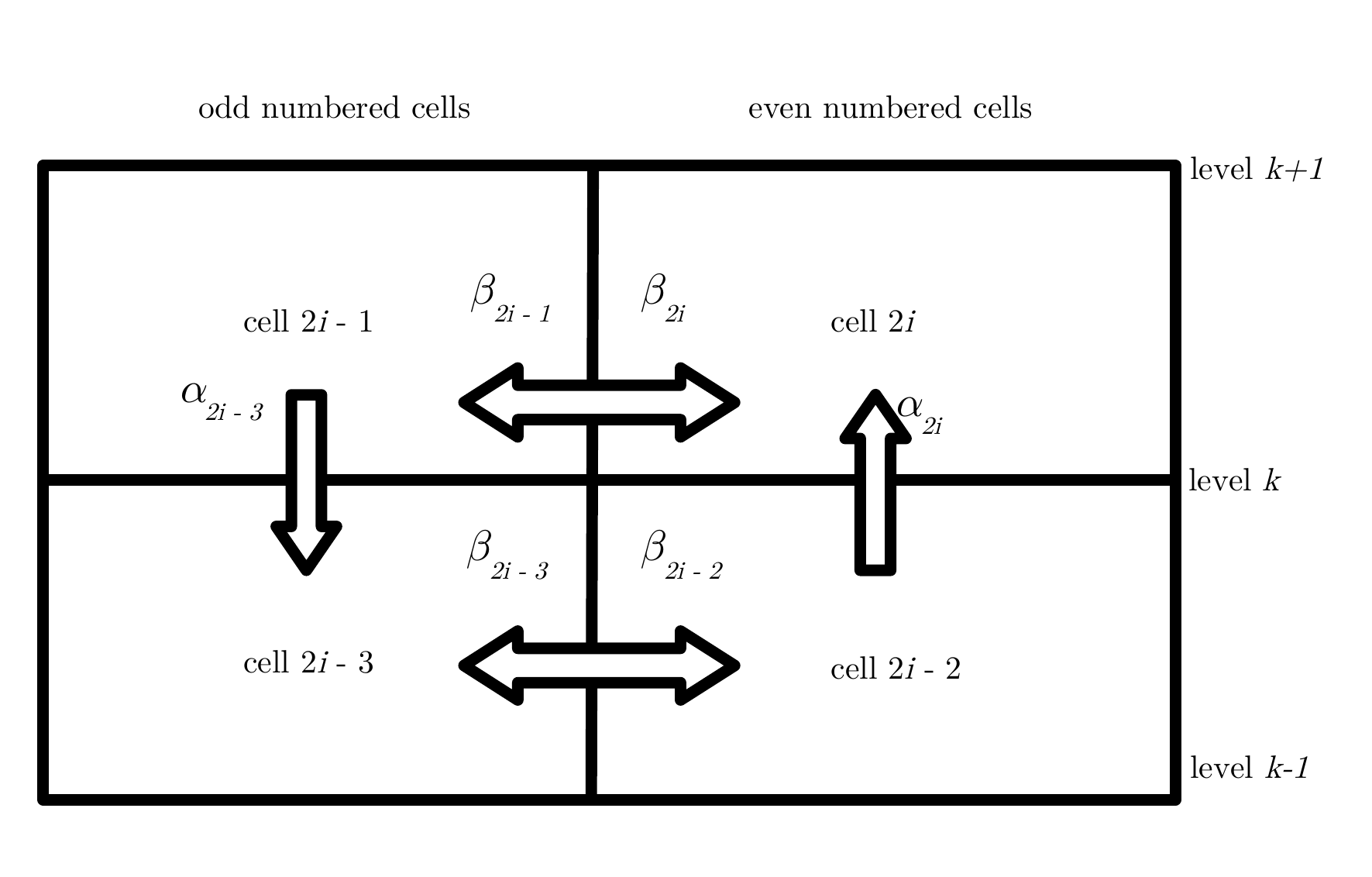}
   \caption{A set of four cells in the up and down streams of the two stream mixing regime.}
   \label{fig:twostreamlevel}
\end{figure}

Our two stream phenomenological advective mixing approach is shown in Fig.~\ref{fig:twostreamlevel} \citep{cannon93}. We set the value of the diffusion coefficient $D_t$ in MONSTAR according to Equation~\ref{eq:Dthm} and calculate $v$ and $l$ for use in MONSOON using
\begin{equation}
\label{eq:length}
l = \alpha H_P,
\end{equation}
where $\alpha = \alpha_{\rm MLT} = 1.69$ and is the same as used in MONSTAR, and from Equation~\ref{eq:diff}
\begin{equation}
\label{eq:velocity}
v = \cfrac{3D}{l}.
\end{equation}
The total mass flux $F$ across radius $r$ is zero in accordance with mass conservation. Using this information the mass fluxes in the up and down streams are calculated using
\begin{equation}
\label{eq:fluxes}
F_u = F_d = \cfrac{4 \pi r^2 \rho v}{2}.
\end{equation}

Fig.~\ref{fig:twostreamlevel} schematically shows the two streams in this mixing regime, with each stream being divided by horizontal levels $k$. The thermohaline mixing velocities according to Equation~\ref{eq:velocity} are calculated at the boundary of each horizontal level. From the structure model, the density and radius are also known at the boundary of each level. To determine the velocity of material in each stream at each level $k$ ($v_{u,k}$ and $v_{d,k}$ for the up and down stream respectively), we denote the fractional cross-sectional areas of the downstream, $f_d$, and the upstream, $f_u$, for the thermohaline region (and all regions where mixing occurs). The values of $f_d$ and $f_u$ also satisfy $f_d + f_u = 1$. To conserve mass it is required that up and down mass fluxes are equal, therefore the flux in each stream is
\begin{equation}
\label{eq:fluxupdown}
4\pi r_k^2 \rho_k v_{u,k}f_u = 4\pi r_k^2 \rho_k v_{d,k}f_d.
\end{equation}
The velocities $v_{u,k}$ and $v_{d,k}$ are determined from the thermohaline velocity in Equation \ref{eq:velocity} by
\begin{equation}
\label{eq:v}
v_{u,k} f_u = v_{d,k} f_d = \cfrac{v}{2}.
\end{equation}
The factor of $\sfrac{1}{2}$ in the right-hand side of Equation~\ref{eq:v} is a direct result of the mass flux equalling zero according to mass conservation. From this condition it is necessary that the flux in each stream must be equal as shown in Equations~\ref{eq:fluxes} and~\ref{eq:fluxupdown}, therefore the total mass flux is split \textit{evenly} between the two streams regardless of the cross-sectional area of each stream. A wider stream will consist of material travelling at a slower velocity and vice versa for a thin stream in accordance with Equation~\ref{eq:v}.

Showing the derivation for the downstream only (similar equations can be derived for the upstream), an equation for $\alpha$, the fraction of mass replaced in a downstream cell moving vertically per second on level $k$, is given by
\begin{equation}
\label{eq:alpha}
\alpha_{2i-1} = \cfrac{1}{2f_d m_{2i-1}} 4\pi r_k^2 v_{d,k} \rho_k,
\end{equation}
where $m_{2i-1}$ is the mass of cell $2i-1$.

The difference between the vertical mass flow into and out of a given cell must move horizontally if mass is to be conserved. This flow rate is denoted by $b$ and shown below for level $k$
\begin{equation}
\label{eq:b}
b_k = \cfrac{1}{2}\left[4\pi(r_{k+1})^2v_{d,k+1}\rho_{k+1} - 4\pi(r_k)^2v_{d,k}\rho_k\right].
\end{equation}

If $b=0$ we still allow for horizontal mass flow. This is calculated using standard mixing length concepts where a blob of material will travel a mixing length $l$ with velocity $v$ before losing its identity. Let $\beta$ be the fraction of mass replaced in each cell moving horizontally per second. The equation for $\beta$ must also take into consideration the mass flow according to $b$ (Equation~\ref{eq:b}). For each cell on level $k$, we define equations for the fraction of mass replaced in an upstream cell per second, say $\beta_{2i}$, and a downstream cell per second, say $\beta_{2i-1}$, in relation to $b$
\begin{equation}
\label{eq:betabneg}
   \left. \begin{aligned}
             \beta_{2i} &= \cfrac{v_k}{f_u l_k}+b_k\\
             \beta_{2i-1} &= \cfrac{v_k}{f_d l_k}
          \end{aligned}
          \right\}
          \qquad \text{if $b_k$ is negative,}
\end{equation}
\begin{equation}
\label{eq:betabpos}
   \left. \begin{aligned}
             \beta_{2i} &= \cfrac{v_k}{f_u l_k}\\
             \beta_{2i-1} &= \cfrac{v_k}{f_d l_k}-b_k
          \end{aligned}
          \right\}
          \qquad \text{if $b_k$ is positive.}
\end{equation}
Equations~\ref{eq:betabneg} and~\ref{eq:betabpos} state that if $b$ is negative, mass must flow from the upstream into the downstream. If $b$ is positive, mass must flow from the downstream into the upstream. We take $f_u = f_d = 0.5$ except in \S\ref{subsec:exp3results}.



\section{Stellar models}
\label{sec:models}

Using MONSTAR and MONSOON we evolve our stellar models from the zero-age main sequence (ZAMS) to the helium flash, which terminates RGB evolution. To allow meaningful comparisons with observations we fit our stellar models to the globular cluster NGC 6397 because it is the only globular cluster to date that has observations for both carbon \citep{briley90} and lithium \citep{lind09} along the RGB. It is imperative to have observations of both elements because carbon and lithium burn at different temperatures and are therefore independent indicators of extra mixing, as stated in \S\ref{sec:intro}.

\subsection{Fitting stellar parameters}
\label{subsec:stellarparams}

We test two metallicities that span the range of observed metallicities for NGC 6397. We test [Fe/H] = $-2.03$ \citep{gratton03} and $-1.82$ \citep{reid98}. The metallicity given in \citet{gratton03} is very close to the \citet[][2010 edition]{harris96} value of $-2.02$. \citet{lind09} find [Fe/H] = $-2.10$, which is similar to results found by \citet{castilho00} ([Fe/H] = $-2.0 \pm 0.05$), \citet{gratton01} ([Fe/H] = $-2.03 \pm 0.02 \pm 0.04$)\footnote{Where the first set of error bars are internal errors and the second set are systematic errors within the abundance scale of the 25 subdwarfs analysed by \citet{gratton01}.}, \citet{korn07} ([Fe/H] = $-2.12 \pm 0.03$), and \citet{husser16} ([Fe/H] = $-2.120 \pm 0.002$). The \citet{reid98} value of $-1.82$ is the upper bound of NGC 6397 published metallicities. 

We use $[\alpha/\text{Fe}]=0.34$ according to \citet{gratton03}. We determine $Z=0.000246$ using [Fe/H] = $-2.0$ according to \citet{gratton03} and $Z=0.00039$ using [Fe/H] = $-1.8$ according to \citet{reid98}.

We construct stellar models with masses $0.79 \text{M}_{\odot}$ and $0.8 \text{M}_{\odot}$ and $Y = 0.24$. These produce main sequence turn-off (MSTO) ages of 13.8 Gyr and 13.2 Gyr respectively, and RGB tip ages of 15.1 Gyr and 14.4 Gyr respectively with [Fe/H] = $-2.0$, which are good matches for the age of NGC 6397 found by \citet{gratton03} of $13.9 \pm 1.1$ Gyr. A $0.8 \text{M}_{\odot}$ stellar model with [Fe/H] = $-1.8$ produces a MSTO age of 13.3 Gyr and an RGB tip age of 14.6 Gyr.

\subsubsection{FDU and RGB bump magnitudes}
\label{subsubsec:fdubump}

The FDU and RGB bump absolute visual magnitudes of NGC 6397 giants are $M_{V} = 3.3$ and 0.16 respectively \citep{lind09} using the bolometric corrections of \citet{alonso99}. \citet{lind09} do not report uncertainties in their magnitude estimates. \citet{nataf13} find the \textit{V}-band magnitude of the bump to be $12.533 \pm 0.046$. We apply this uncertainty to the bump magnitude of \citet{lind09} and find the values of the FDU and bump magnitudes to be $3.3 \pm 0.046$ and $0.16 \pm 0.046$ respectively.

\begin{figure}
   \centering
   \includegraphics[width=\columnwidth]{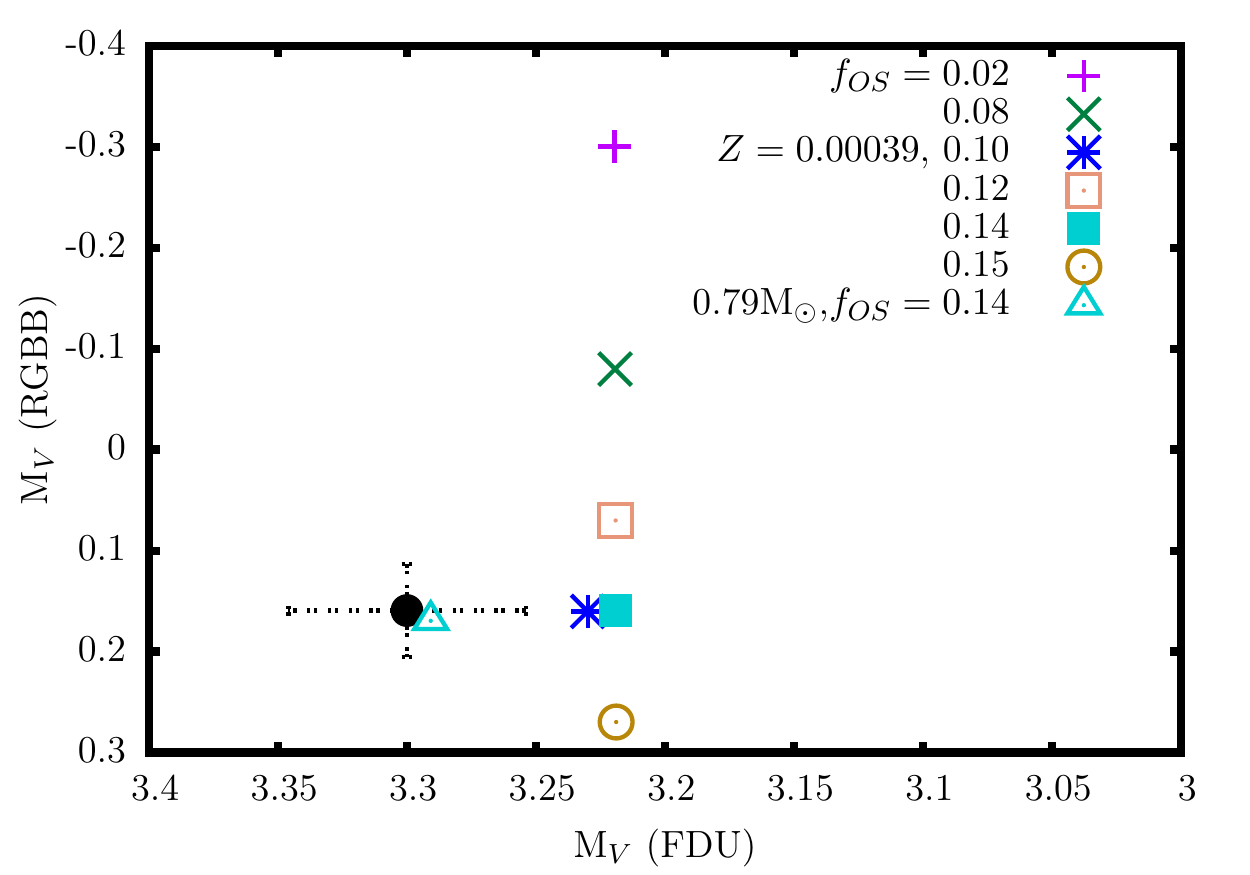}
   \caption{FDU and RGB bump magnitudes for our $0.8 \text{M}_{\odot}$, $Z=0.000246$, $\text{[Fe/H]} = -2.0$ stellar model where we vary $f_{OS}$, with values given in the legend. Also given in the legend are the best-fit models of $0.8 \text{M}_{\odot}$, $Z=0.00039$, $\text{[Fe/H]} = -1.8$ and $f_{OS} = 0.10$ (labelled with $Z$), and $0.79 \text{M}_{\odot}$, $Z=0.000246$, $\text{[Fe/H]} = -2.0$ and $f_{OS} = 0.14$. FDU and RGB bump magnitudes of \citet{lind09} with the observational uncertainty of \citet{nataf13} is black.}
   \label{fig:fdubump}
\end{figure}

As stated in \S\ref{subsubsec:overshoot}, the value of the overshoot factor $f_{OS}$ affects the predicted magnitude of the bump. We test a range of $f_{OS}$ values from 0.02 to 0.15 to match the magnitudes of FDU and RGB bump, with results shown in Fig.~\ref{fig:fdubump} for our $0.8 \text{M}_{\odot}$, $Z=0.000246$, $\text{[Fe/H]} = -2.0$ stellar model.

We find the stellar models that best match the FDU and bump magnitudes observed for NGC 6397 using our two chosen metallicities are a $0.79 \text{M}_{\odot}$ star with $Z=0.000246$ using [Fe/H] = $-2.0$ \citep{gratton03}, $Y=0.24$ and $f_{OS}=0.14$ (henceforth our $0.79 \text{M}_{\odot}$ model), and a $0.80 \text{M}_{\odot}$ star with $Z=0.00039$ using [Fe/H] = $-1.8$ \citep{gratton03}, $Y=0.24$ and $f_{OS}=0.10$ (henceforth our $0.80 \text{M}_{\odot}$ model).

\subsubsection{Hertzsprung-Russell diagram}
\label{subsubsec:hrd}

\begin{figure}
   \centering
   \includegraphics[width=\columnwidth]{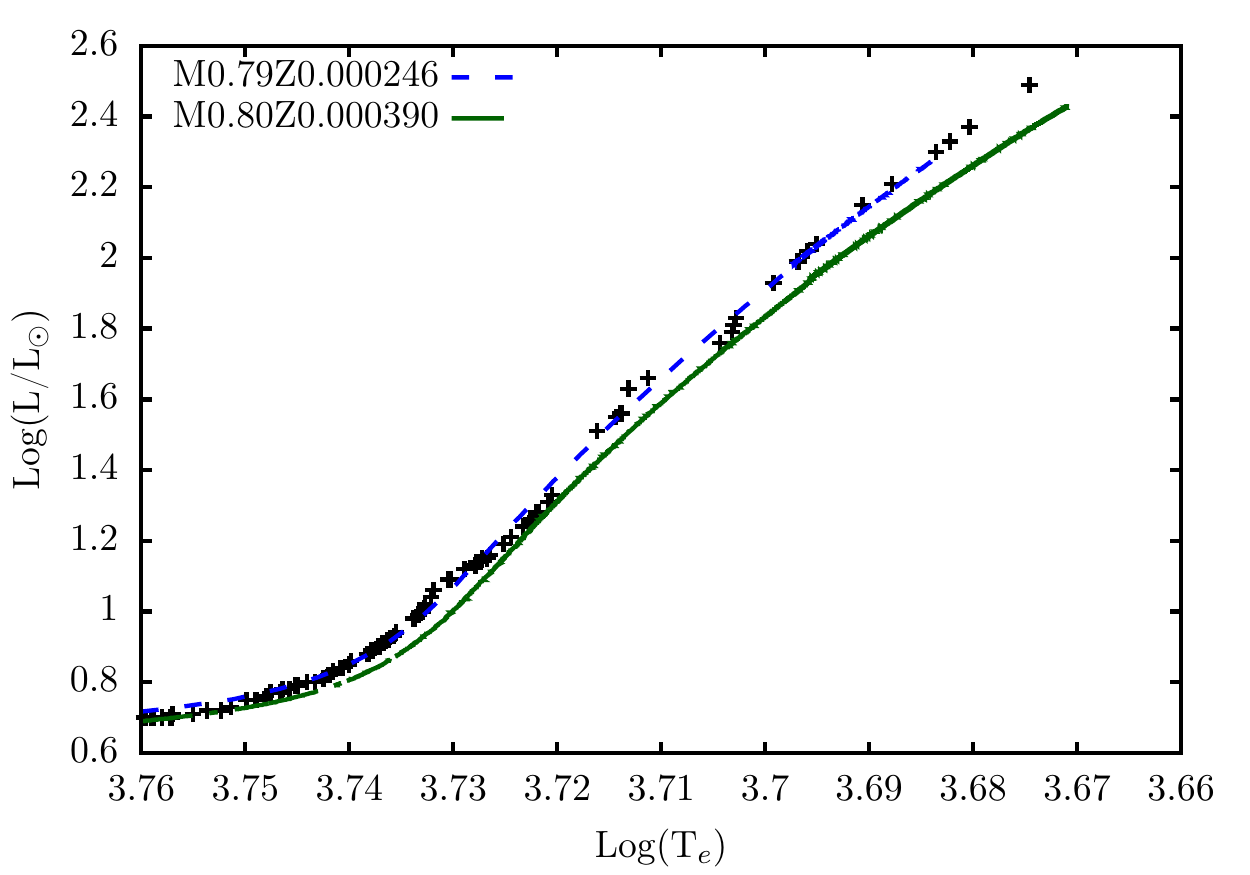}
   \caption{The giant branch of the Hertzsprung-Russell diagram of our best-fit models compared to observations of NGC 6397 \citep{lind09}. Our $0.79 \text{M}_{\odot}$ model, denoted M0.79Z0.000246 in the legend, is the blue dashed curve. Our $0.80 \text{M}_{\odot}$ model, denoted M0.8Z0.00390, is the green solid curve.}
   \label{fig:hrd}
\end{figure}

We plot the giant branch of our best-fit models according to our overshoot results (Fig.~\ref{fig:fdubump}) on a HRD in Fig.~\ref{fig:hrd} compared to the observational HRD of NGC 6397 \citep{lind09}.

In Fig.~\ref{fig:hrd} we see that the model that best matches the giant branch of NGC 6397 is our $0.79 \text{M}_{\odot}$ model. This is also the best fit for FDU and the LFB as discussed in \S\ref{subsubsec:fdubump}. The difference in metallicities between our $0.79 \text{M}_{\odot}$ model and $0.80 \text{M}_{\odot}$ model (along with initial stellar mass) likely contributes to the HRD discrepancies in Fig.~\ref{fig:hrd}, though this does not change our results or conclusions.

\begin{figure}
   \centering
   \includegraphics[width=\columnwidth]{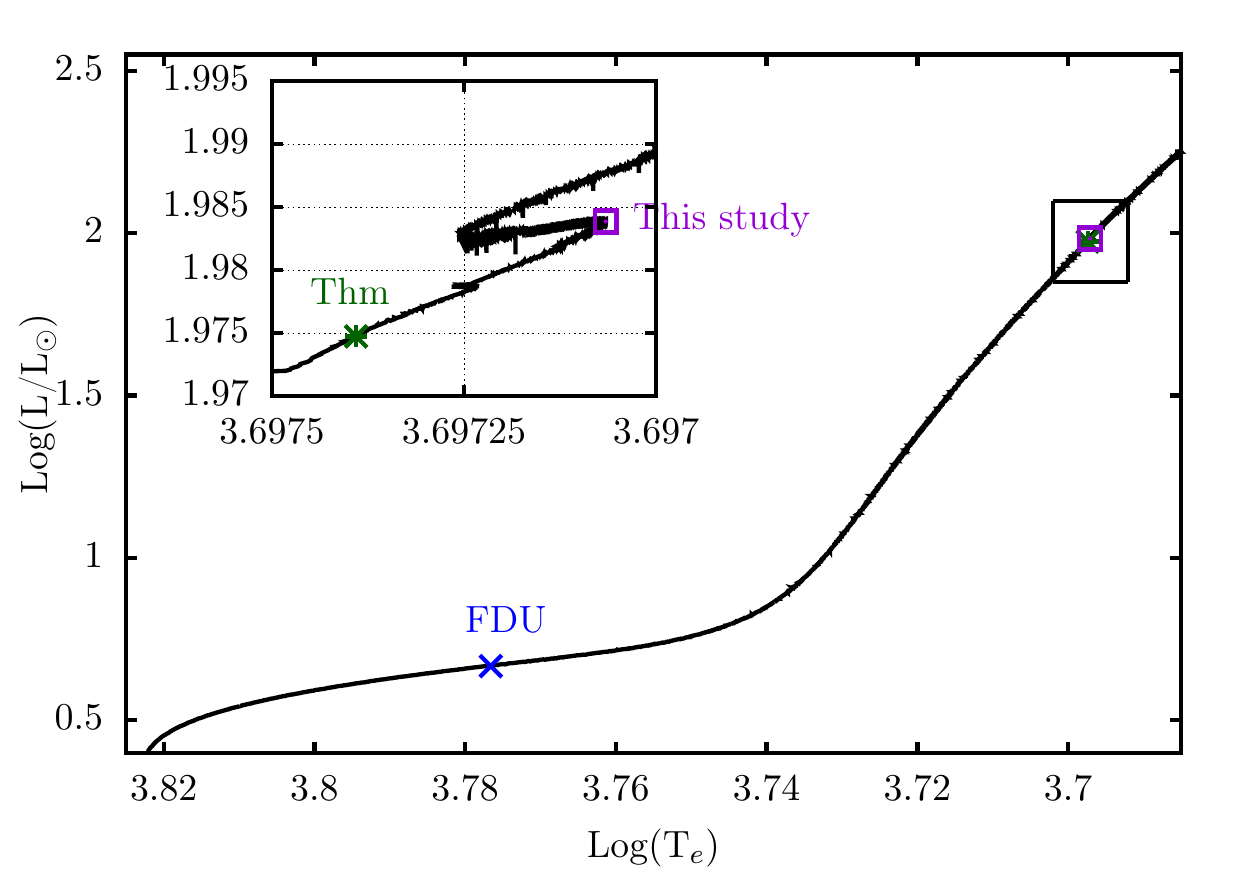}
   \caption{Hertzsprung-Russell diagram of our $0.79 \text{M}_{\odot}$ model. The points on the plot denote the location on the HRD where abundance changes occur in our models. First dredge-up (``FDU'', blue cross), thermohaline mixing (``Thm'', green asterisk), and the luminosity we have chosen to take snapshots of our mixing mechanism (``This study'', purple square) are indicated. The region of the LFB is shown in the inset.}
   \label{fig:model6000}
\end{figure}

To allow analysis of burning timescales and elemental changes in our model, we choose an epoch just after surface abundance changes due to thermohaline mixing have occurred. The epoch chosen is shown in Fig.~\ref{fig:model6000}. All analysis of internal stellar properties is done at this epoch.

\begin{figure}
   \centering
   \includegraphics[width=\columnwidth]{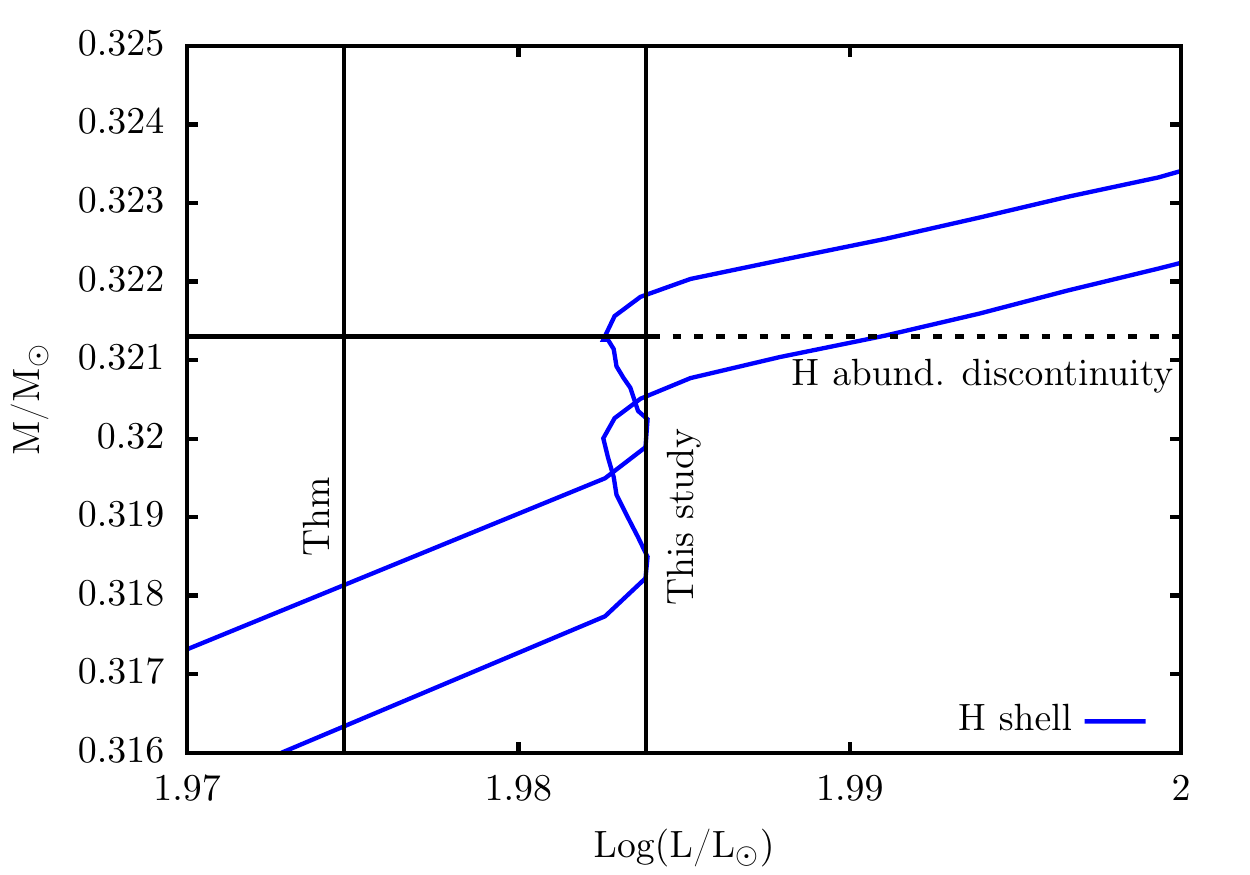}
   \caption{Upper and lower mass boundaries of the H shell (in blue) of our $0.79 \text{M}_{\odot}$ model as it is nearing the LFB. These mass boundaries are defined as the mass where X = 0.05 and X = 0.66. The luminosity location of surface abundance changes due to thermohaline mixing (``Thm'') and the epoch chosen for this study (``This study'') are indicated and are the same as in Fig.~\ref{fig:model6000}. The initial mass location of the hydrogen abundance discontinuity is also shown as a solid line until it is erased at the LFB where it becomes a dashed line.}
   \label{fig:thermo}
\end{figure}

The locations of FDU and thermohaline mixing (``Thm'') on the HRD are when abundance changes on the surface are observed due to these events. Interestingly we see from the inset of Fig.~\ref{fig:model6000} that abundance changes due to thermohaline mixing first appear on the surface at a lower luminosity than the LFB. Generally it is thought that thermohaline mixing coincides with the LFB on the HRD but this is not necessarily the case according to Fig.~\ref{fig:model6000} and is also found by other groups\footnote{As shown in Fig.~\ref{fig:model6000} the difference in $\log(\text{L}/\text{L}_{\odot})$ between thermohaline mixing and the LFB is only $\sim 0.01$. Observationally this is extremely difficult to distinguish, hence it is sufficient to approximate that the LFB and extra mixing occur at the same time.} \citep{charbonnel10}. Extra mixing is governed by $^3$He fusion inverting the $\mu$ gradient whereas the LFB is due to the structural change in the star as the H shell passes through the hydrogen discontinuity caused by FDU. These events need not coincide with each other because they occur at different temperatures. The lithium abundance changes at lower temperatures before the H shell reaches the discontinuity. Hence abundance changes caused by thermohaline mixing occur slightly before the reversal in the HRD, which requires the H shell to have reached the discontinuity. There is also a delay between when thermohaline mixing begins and when it connects to the convective envelope as Fig~\ref{fig:thermo} shows in more detail. The amount of time elapsed is dependent upon timestepping and spatial resolution in the stellar evolution code \citep[for a detailed discussion of this, see][]{lattanzio15}. In Fig.~\ref{fig:thermo} we can see that thermohaline mixing begins prior to the H shell connecting with the hydrogen abundance discontinuity.

\subsection{Nuclear burning timescales in our standard model}
\label{subsec:burningtimescales}

The timescales discussed below in \S\ref{subsubsec:libe} have been determined using MONSTAR. We denote $D_0$ as the thermohaline coefficient for the standard case with $C_t = 1000$ and test three variations of $D_t$ according to $D_t = f \times D_{0}$ where $f = \sfrac{1}{3}$, 1, 3. We show the results in Figs.~\ref{fig:taucombined} and~\ref{fig:Xelements}. The case of $f = 3$ is mixing that is faster than the standard $f = 1$ case, and $f = \sfrac{1}{3}$ is mixing that is slower.

The mixing timescales for the three cases ($f$ = $\sfrac{1}{3}$, 1, 3) and various nuclear burning timescales are shown in Fig.~\ref{fig:taucombined}. Timescales are plotted as a function of radius. The nuclear burning timescales depend upon temperature and density, and remain the same regardless of $f$.

\subsubsection{Lithium and carbon burning timescales in the thermohaline region}
\label{subsubsec:libe}

\begin{figure}
   \centering
   \includegraphics[width=\columnwidth]{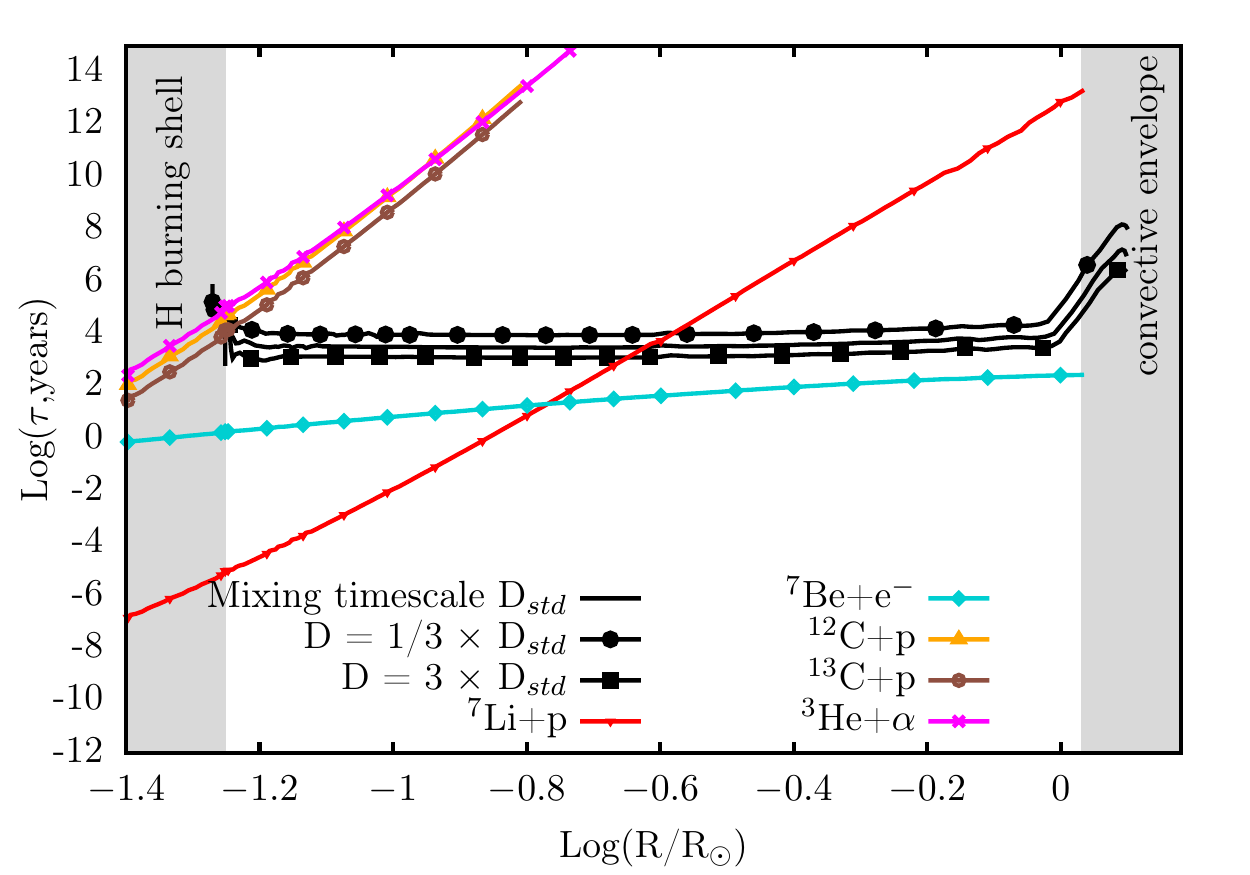}
   \caption{The log of mixing timescales (years; black curves) and nuclear burning timescales of $^7$Li (red), $^7$Be (turquoise), $^{12}$C (orange), $^{13}$C (burgundy), and $^3$He (magenta) in our $0.79 \text{M}_{\odot}$ model at the point labelled ``This study'' in Fig.~\ref{fig:model6000}.}
   \label{fig:taucombined}
\end{figure}

\begin{figure}
   \centering
      \includegraphics[width=\columnwidth]{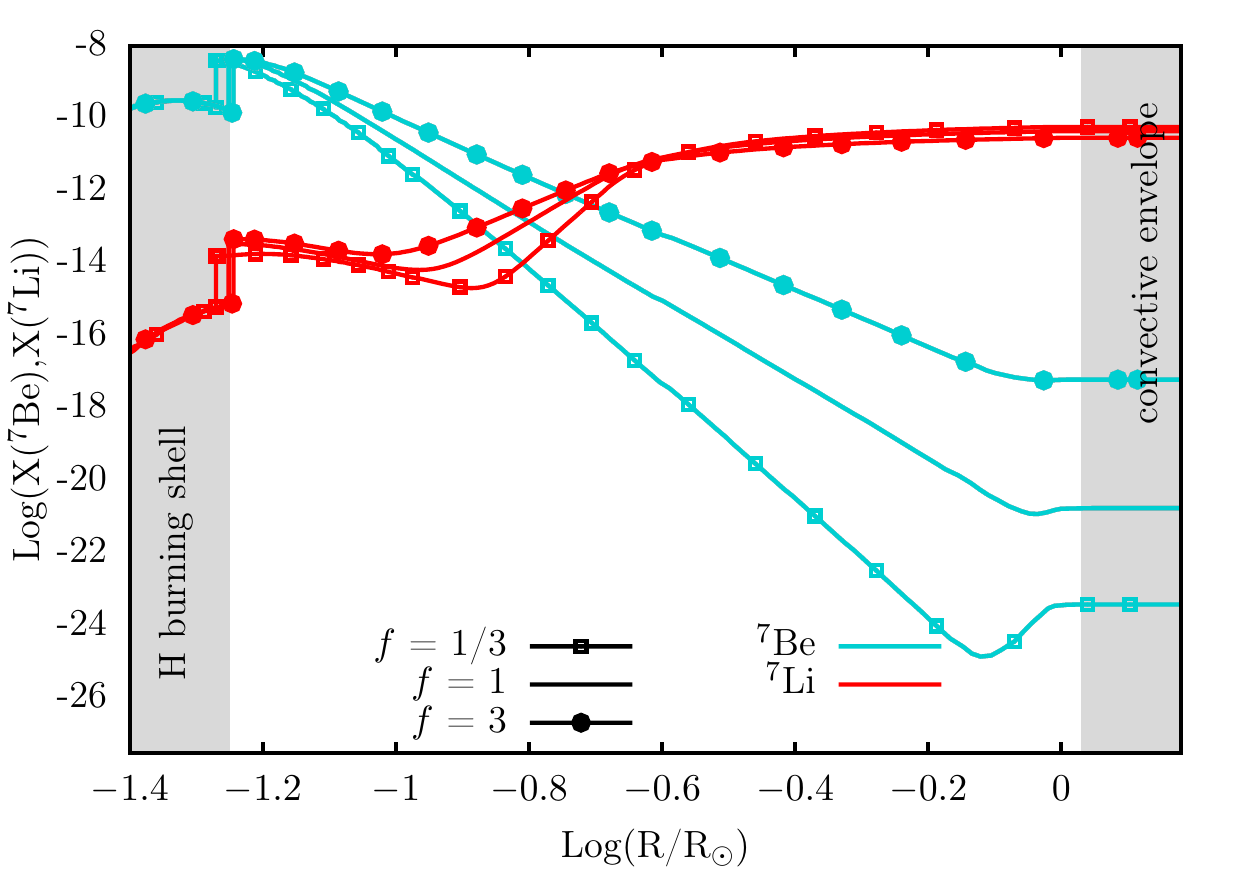}
      \includegraphics[width=\columnwidth]{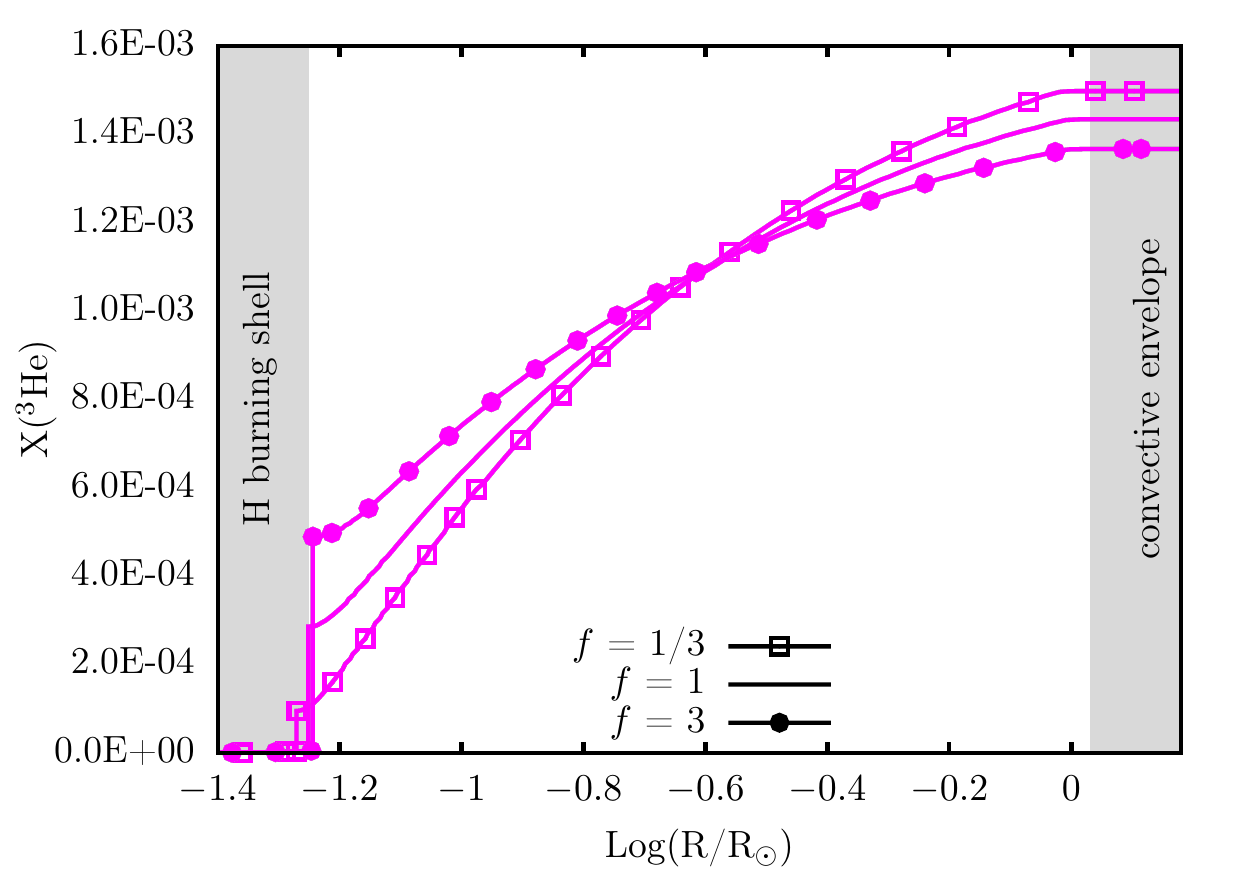}
      \includegraphics[width=\columnwidth]{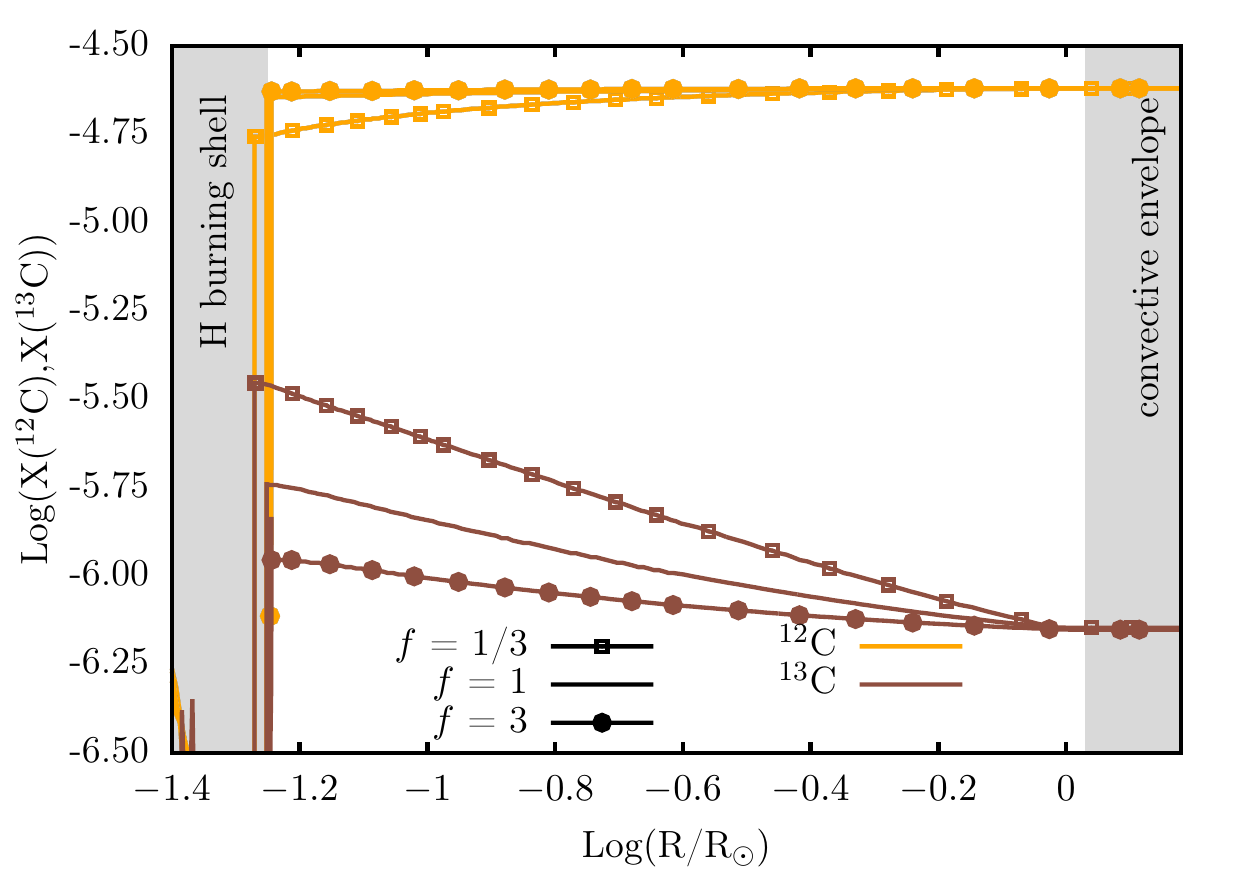}
   \caption{Abundance profiles for $f$ = $\sfrac{1}{3}$ (open squares), 1 (solid curves), and 3 (filled circles). Top panel: $^7$Li (red) and $^7$Be (turquoise). Middle panel: $^3$He. Bottom panel: $^{12}$C (orange) and $^{13}$C (burgundy) in our $0.79 \text{M}_{\odot}$ model at the point labelled ``This study'' in Fig.~\ref{fig:model6000}.}
   \label{fig:Xelements}
\end{figure}

\begin{figure}
   \centering
   \includegraphics[width=\columnwidth]{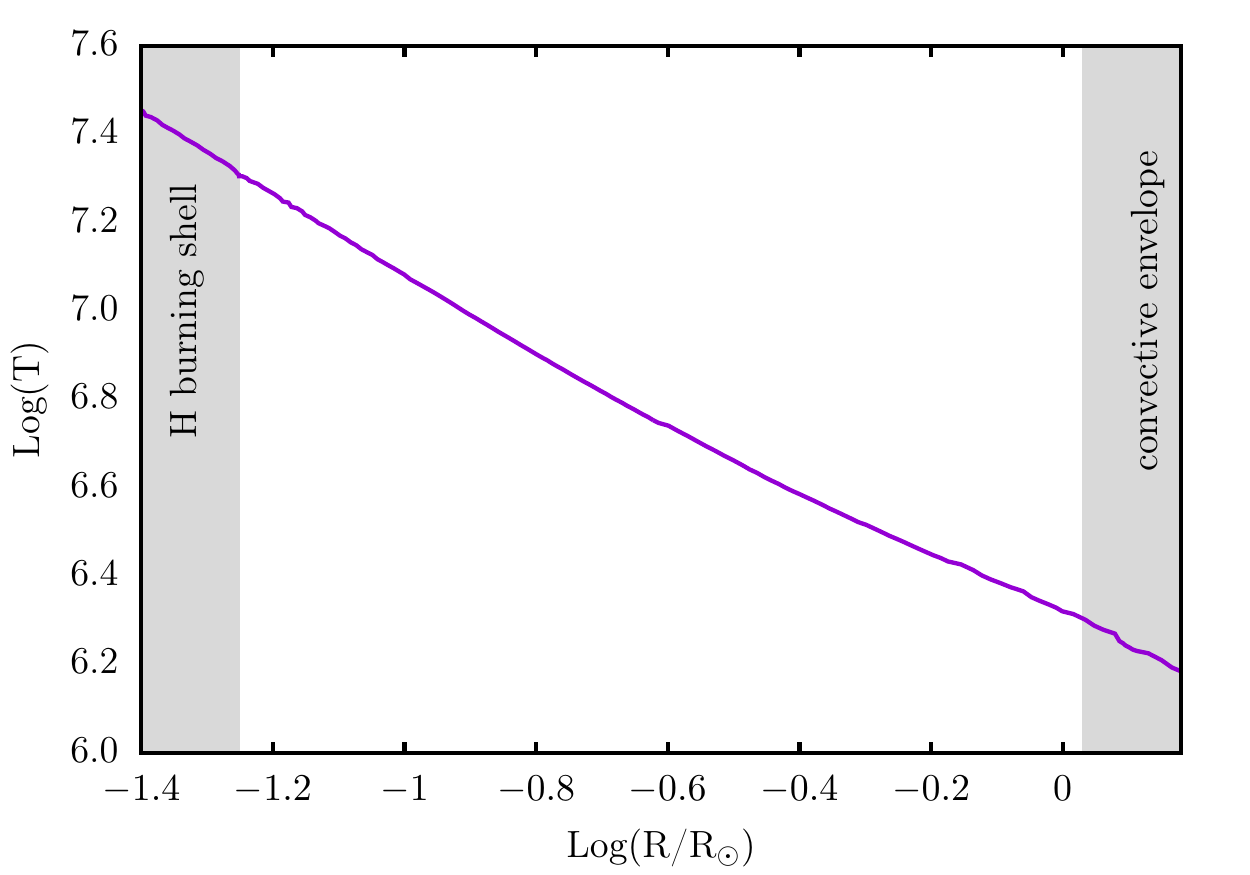}
   \caption{Temperature as a function of radius in the thermohaline region for our $0.79 \text{M}_{\odot}$ stellar model. The stage of evolution is shown in Fig.~\ref{fig:model6000} as the point ``This study''.}
   \label{fig:temp}
\end{figure}

There are two limiting cases for the mixing: if it is infinitely fast then all abundances are homogeneous; if there is no mixing the shapes of the abundance profiles are due to burning alone (decreasing with increasing depth/temperature). Although not dominant at the temperatures in the thermohaline region\footnote{At a time just after the initiation of thermohaline mixing, temperatures range from $\sim 20$ MK at the base of the thermohaline region to $\sim 2$ MK at the base of the convective envelope (Fig.~\ref{fig:temp}).} as shown in Fig.~\ref{fig:temp}, the ppII chain is still operating, destroying $^3$He by alpha capture via $^{3}\text{He}(\alpha,\gamma)^{7}\text{Be}$ and producing $^7$Be. As stated previously, faster mixing ($f = 3$ in Figs.~\ref{fig:taucombined} and~\ref{fig:Xelements}) produces a more homogeneous $^3$He profile than slower mixing ($f = \sfrac{1}{3}$). The timescale for $^3$He alpha capture is longer than the mixing timescale regardless of $f$, producing a $^3$He abundance that does not vary dramatically with $f$ as shown in the middle panel of Fig.~\ref{fig:Xelements}.

Alpha capture on $^3$He produces $^7$Be and this occurs predominantly in the interior where $^3$He burns fastest. A fast rate of mixing will bring $^3$He from the interior (where temperatures are higher) towards the surface faster. Therefore less $^3$He will burn, producing less $^7$Be. However, the $^7$Be will also be brought from the interior faster, and although it will be destroyed via electron capture at nearly the same rate at all temperatures, the amount of $^7$Be destroyed is less than when mixing is slower.

The rate of $^7$Be destruction is always faster than the rate of mixing for all tested $f$, therefore over the entire thermohaline zone the $^7$Be abundance does not become homogeneous. In the interior (from the base of the thermohaline zone to $\log(\text{R}/\text{R}_{\odot}) \sim -0.8$), $^7$Li destruction is much faster than mixing by several orders of magnitude, therefore the $^7$Li profile is not homogeneous with position. The $^7$Li profile is complicated because both destruction (by proton capture) and production (by electron capture on $^7$Be) timescales are of similar orders of magnitude, particularly around $\log(\text{R}/\text{R}_{\odot}) \sim -0.8$. This produces the variation of around 3 orders of magnitude that is dependent upon $f$ and shown in the top panel of Fig.~\ref{fig:Xelements}. Beyond $\log(\text{R}/\text{R}_{\odot}) \sim -0.8$ the rate of $^7$Li destruction is slower than the rate of mixing and the profile is approximately homogeneous.

Faster mixing means that the timescale of $^{12}$C destruction is slower than the mixing timescale over the entire thermohaline region and the abundance profile is (almost) homogeneous. When $f$ is reduced and the rate of mixing is slower, the timescale of $^{12}$C destruction can be faster than the mixing timescale for a thin region at the base of the thermohaline zone. More $^{12}$C destruction occurs because $^{12}$C is allowed to burn at the base of the thermohaline zone.

Destruction of $^{13}$C via proton capture is slower than both the rates of mixing and $^{12}$C destruction. When mixing is fast, $^{13}$C is essentially homogenised over the region. When mixing is slower, $^{13}$C is destroyed slower than it is produced, resulting in net production of $^{13}$C. The combination of increased $^{12}$C destruction and increased $^{13}$C production results in a $^{12}$C abundance that is lower and a $^{13}$C abundance that is higher in the thermohaline region when $f = \sfrac{1}{3}$ compared to the $f = 1$ and $f = 3$ cases.

\subsection{Motivation for studies and observational limitations}
\label{subsec:standardmodels}

The data we use for NGC 6397 are from \citet{lind09} for lithium and \citet{briley90} for carbon. \citet{angelou15} found that the decrease in the carbon abundances due to thermohaline mixing occurred prior to the RGB bump whereas lithium declined at the bump, as predicted. We have found that this is related to the distance modulus and note that \citet{lind09} and \citet{briley90} used different values of the distance modulus for NGC 6397; \citet{lind09} report 12.57 and \citet{briley90} report 11.8. By adjusting the distance modulus of the stars observed by \citet{briley90} to match that of \citet{lind09}, we resolve the issue found by \citet{angelou15} and find that carbon and lithium begin their decrease at the same magnitude.

\begin{figure}
   \centering
      \includegraphics[width=\columnwidth]{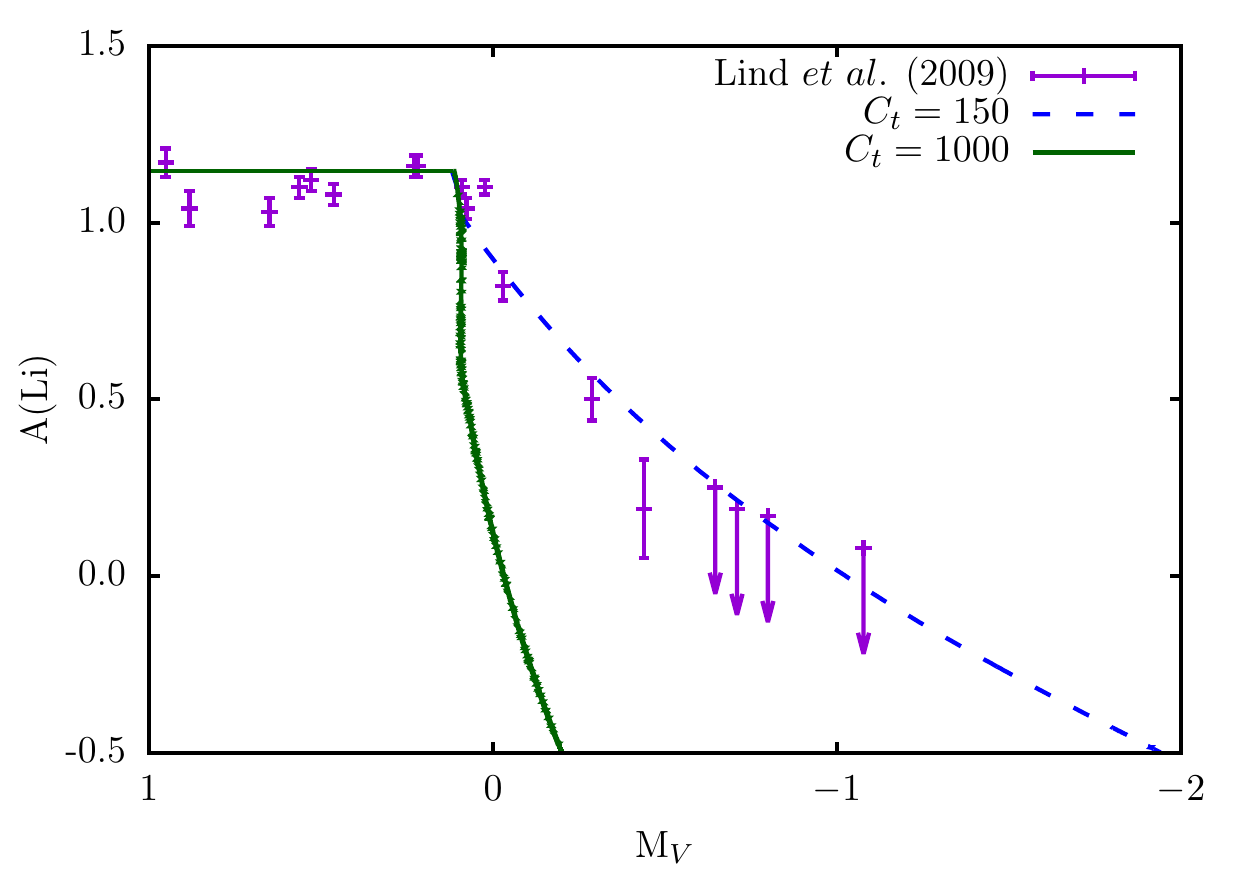}
      \includegraphics[width=\columnwidth]{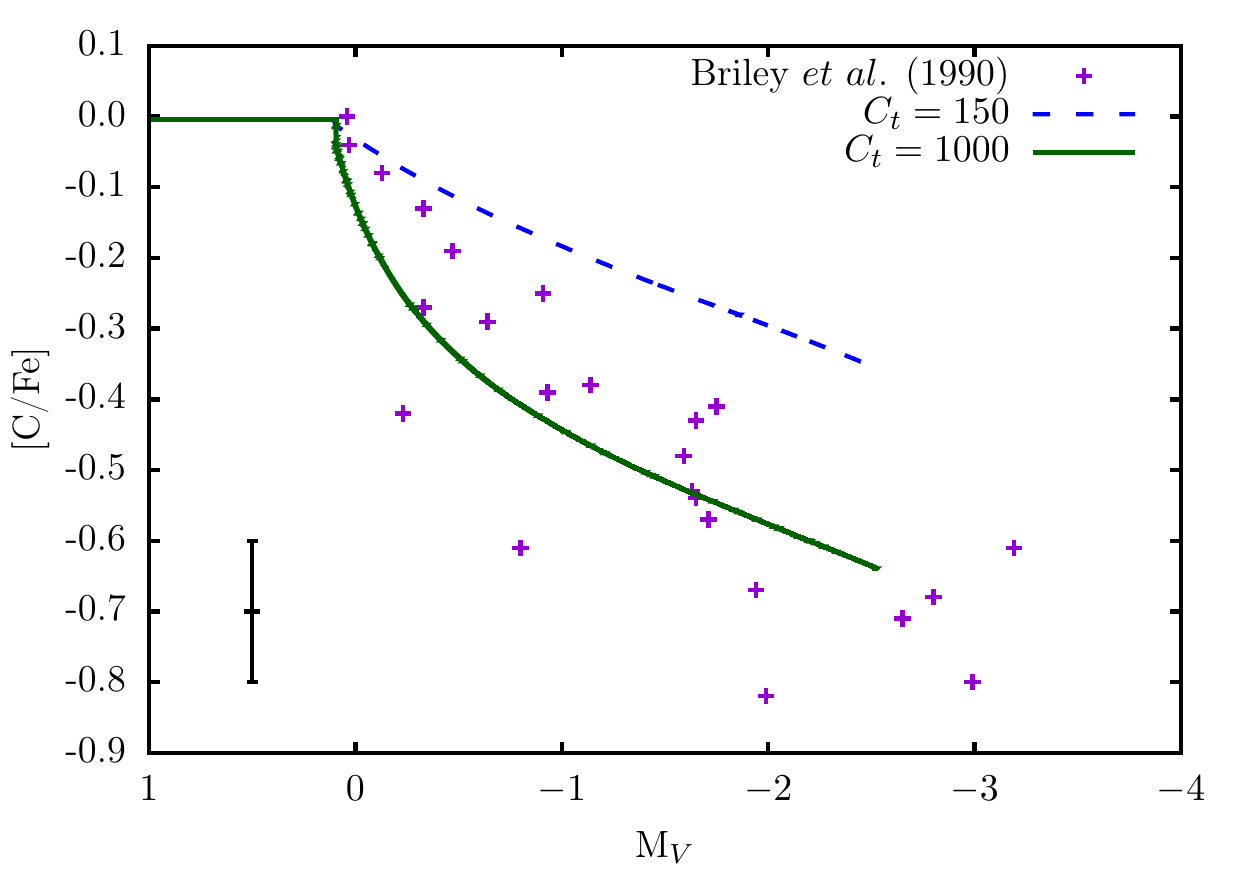}
   \caption{Top panel: Theoretical surface lithium abundances (curves) compared to observations \citep[purple points, arrows indicate upper limits,][]{lind09}. Bottom panel: Theoretical carbon abundances (curves) compared to observations with a 0.1 dex error bar \citep[purple points,][]{briley90}. In each panel the dark green curve is when the thermohaline mixing free parameter $C_t = 1000$. The blue dashed curve is for $C_t = 150$.}
   \label{fig:motivation2}
\end{figure}

To match lithium abundances to observations, a $C_t$ value of around 150 is required but a value closer to 1000 is needed to match carbon as shown in Fig.~\ref{fig:motivation2} \citep[as also found by][]{angelou15}. To elaborate on this, if $C_t = 1000$ is used and carbon matches observations then the models deplete lithium too much. Converse to this, if we use $C_t = 150$ to match lithium observations then we require an increased depletion of carbon.

The spread of the observational data is a complication when comparing to our theoretical models (Fig.~\ref{fig:motivation2}). The spread of the carbon abundances is of the order of 0.2 dex at a given magnitude.  
This spread is most likely due to the different stellar populations present in NGC 6397. 
\citet{angelou12} showed that observations of carbon and nitrogen in the intermediate-metallicity clusters M3 and M13 were best represented by separate models with initially different carbon and nitrogen compositions. The stars with an essentially  normal composition are the CN-weak population, which form the upper envelope of the carbon distribution. It is these stars that we try to fit here. Errors of individual star carbon abundances are around 0.1 dex \citep{briley90}, as shown in the bottom panel of Fig.~\ref{fig:motivation2}. The 0.1 dex error we use for the carbon observations is based upon the discussion and values given (in their Table 11) in \citet{briley90} and is a conservative value for the upper-RGB stars. In fact, the errors given in \citet{briley90} show their coolest stars towards the tip of the RGB have errors larger than their hotter stars, which makes sense given that the strength of the carbon molecular bands decreases with temperature. The model with $C_t = 150$ can match the lower RGB stars within errors but cannot match the most carbon-depleted upper RGB stars as shown in Fig.~\ref{fig:motivation2}.

The observations of lithium are more tightly constrained with smaller errors \citep{lind09} but again there are very few observations near the RGB tip and the coolest few stars do not have associated errors\footnote{The observed abundances for the coolest RGB stars are upper limits only.} as shown in the top panel of Fig.~\ref{fig:motivation2}. When $C_t = 150$ we can match the lithium abundances of all of the RGB stars in NGC 6397 but when $C_t = 1000$ we cannot match any RGB stars after the start of extra mixing.

\section{Test Cases and Results}
\label{sec:results}

\subsection{Case 1: Independently changing $v$ and $l$ in the two stream advective mixing scheme}
\label{subsec:exp1results}

We explore the advective two stream model implemented in MONSOON by varying $v$ and $l$ in Equation~\ref{eq:diff}. We invoke ``mixing factors'' to modify the mixing length and velocity according to
\begin{equation}
\begin{aligned}
\label{eq:vlnew}
v_{\rm new} &= f_v \times v_{\rm std},\\
l_{\rm new} &= f_l \times l_{\rm std},
\end{aligned}
\end{equation}
where the subscript ``std'' indicates the parameter value for a given $C_t$ (as defined by Equation~\ref{eq:Cthm}). Looking at the value of $\beta$ for horizontal mixing in Equations~\ref{eq:betabneg} and~\ref{eq:betabpos}, we see that the only place in our two stream mixing model where $l$ appears is in the equations. Therefore it is straightforward to show that $\beta \propto 1 / l$. Hence varying $f_l$ is the same as varying $1/\beta$.

It is important to note that when $f_v$ and $f_l$ are varied independently we change the effective value of $D_t$ in MONSOON by a factor of $f_v$ or $f_l$ despite setting a value of $D_t$ in MONSTAR according to the Ulrich/Kippenhahn formula. Therefore $D_t$ in MONSTAR and $D_t$ in MONSOON will be inconsistent. Indeed, as soon as we use a different mixing algorithm (in MONSOON) to the diffusion equation (in MONSTAR) then the results are technically inconsistent. We accept this inconsistency because there is negligible feedback on the underlying stellar structure.

Table~\ref{table:exp1} summarises the models calculated for Case 1. When referring to specific tests in Table~\ref{table:exp1} we first refer to the case letter and then the value of the variable. For example, when referencing the test where $f_v$ is being changed independently and we wish to refer to the $f_v = 0.10$ test, we say $V0.10$.

\subsubsection{Single parameter results}
\label{subsubsec:singleparm}

\begin{figure}
   \centering
      \includegraphics[width=\columnwidth]{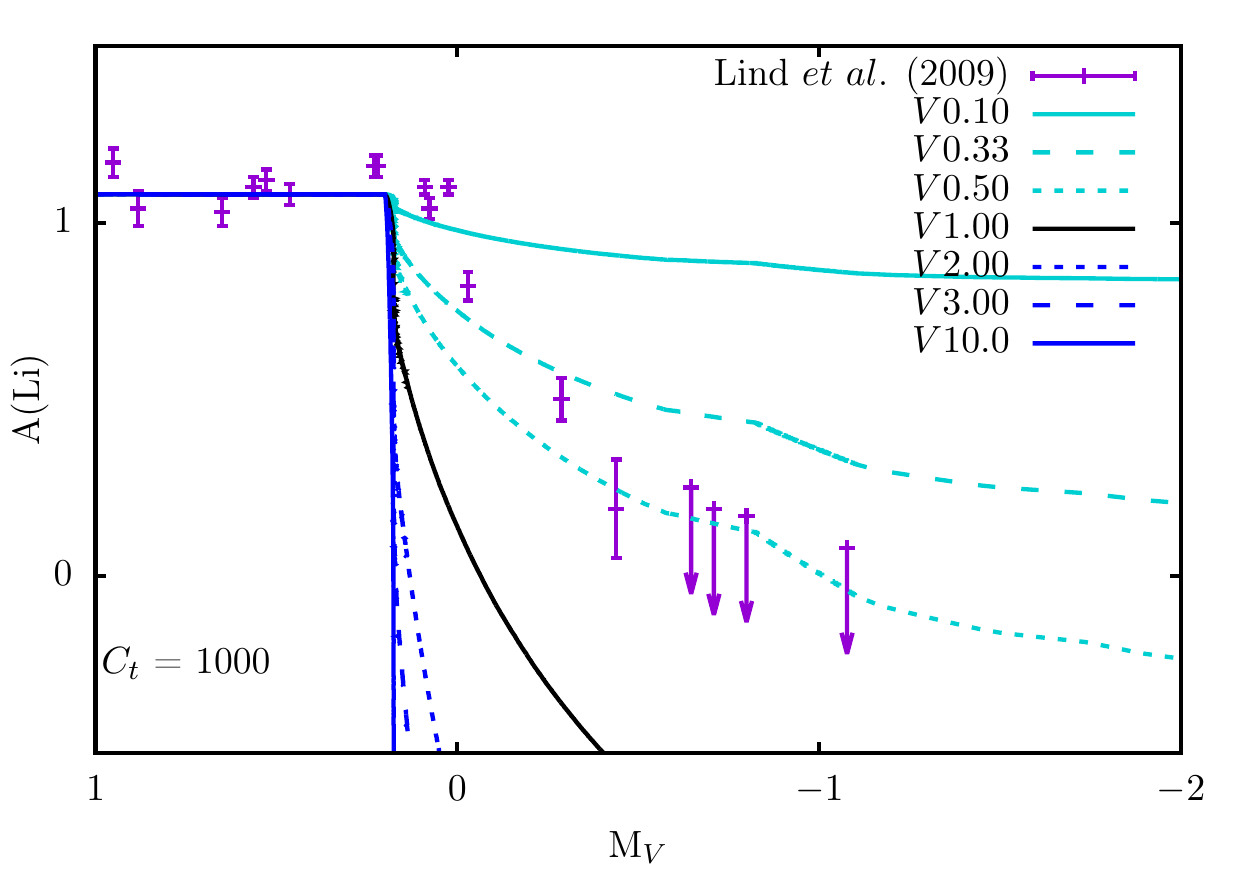}
      \includegraphics[width=\columnwidth]{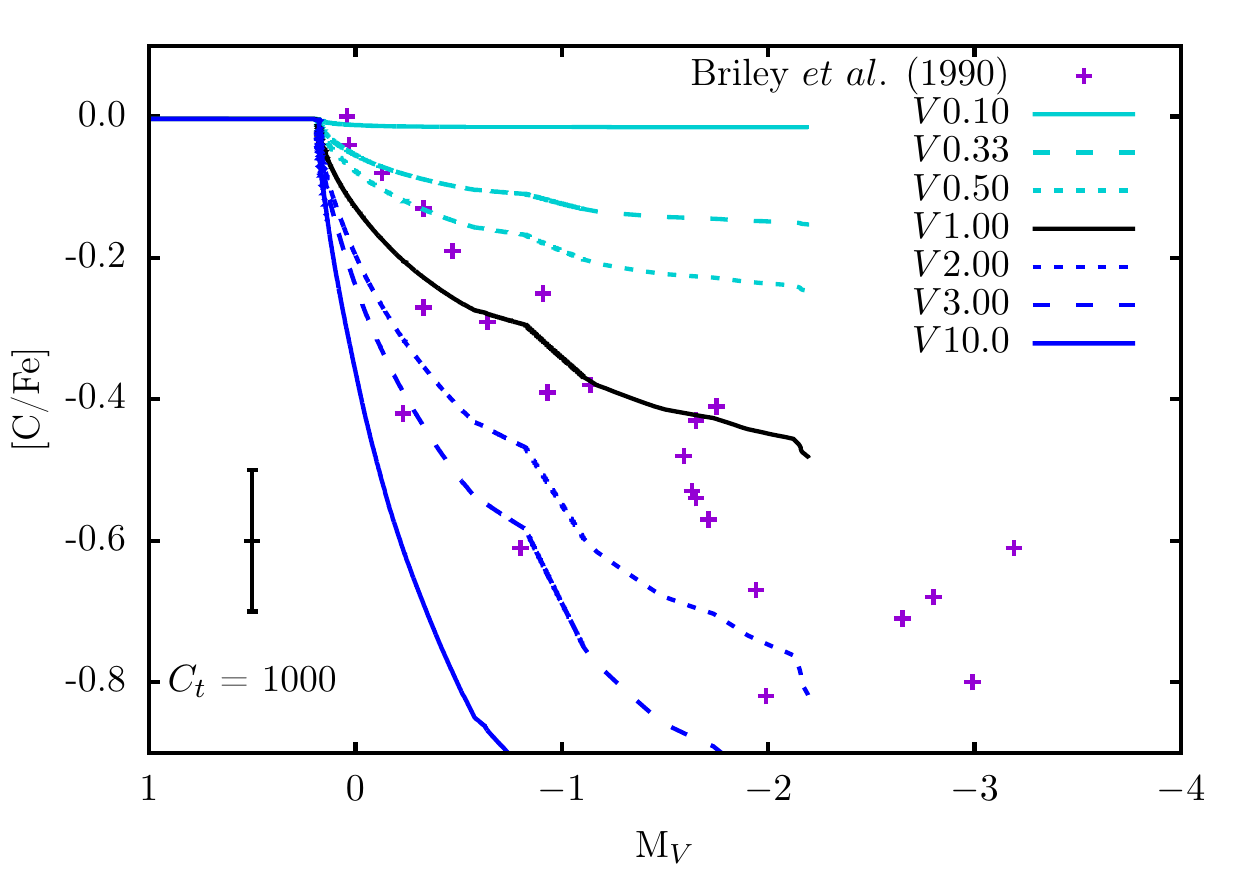}
   \caption{The effect of changing $f_v$ on the surface abundance profiles of A(Li) (top panel) and [C/Fe] with a 0.1 dex error bar (bottom panel) and $C_t = 1000$. The values of $f_v$ and colour key of the curves are in the legend of the plots. Observations are purple points, with upper limits denoted by arrows \citep{briley90,lind09}.}
   \label{fig:fv}
\end{figure}

\begin{figure}
   \centering
      \includegraphics[width=\columnwidth]{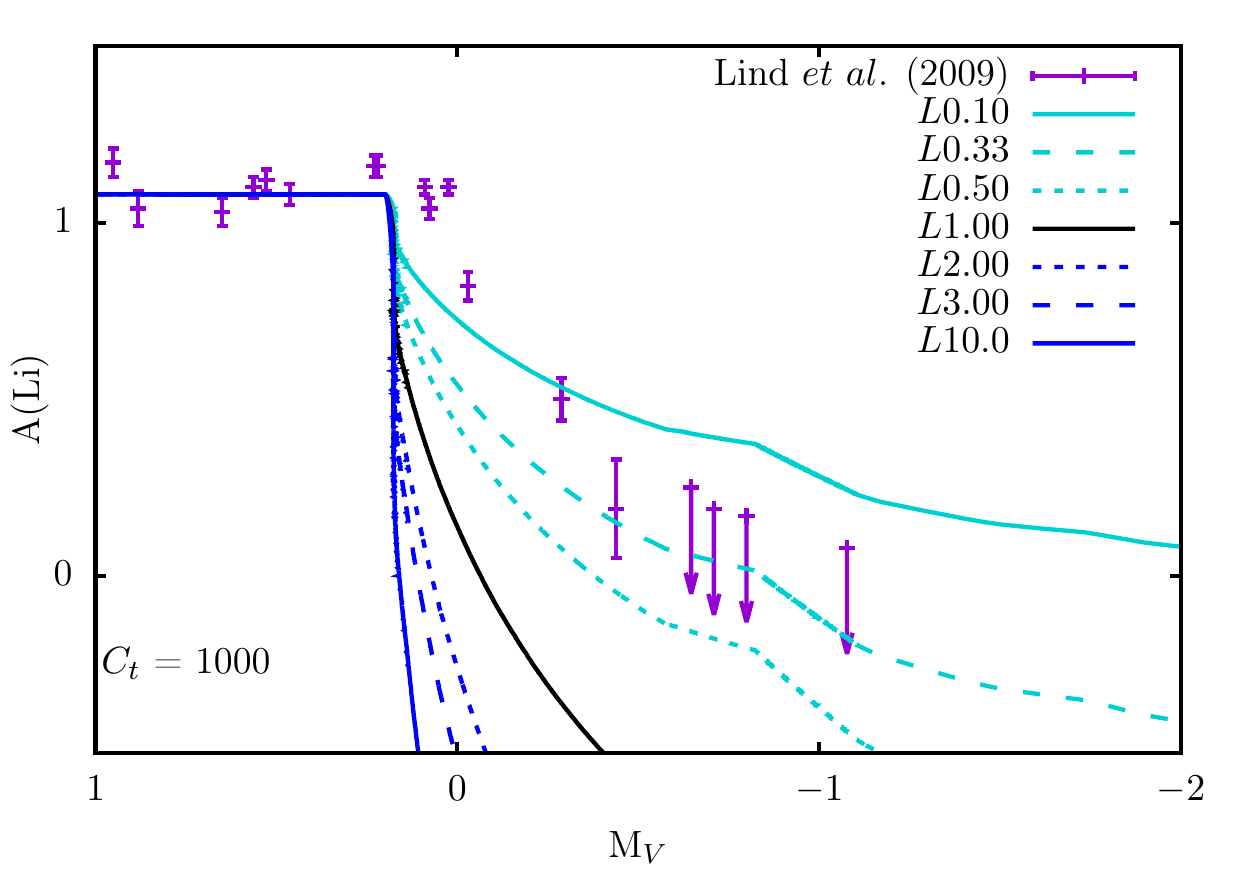}
      \includegraphics[width=\columnwidth]{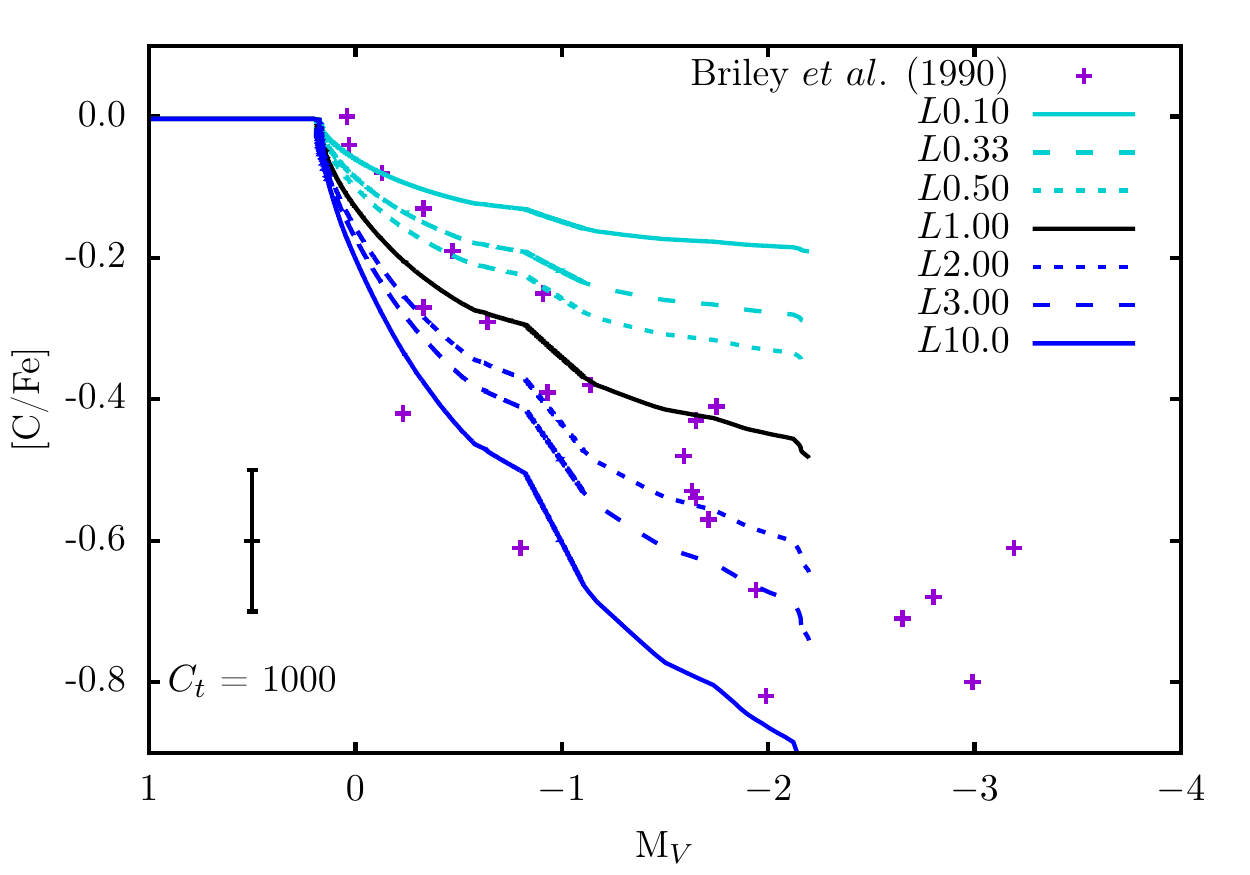}
   \caption{The effect of changing $f_l$ on the surface abundance profiles of A(Li) (top panel) and [C/Fe] with a 0.1 dex error bar (bottom panel) and $C_t = 1000$. The values of $f_l$ and colour key of the curves are in the legend of the plots. Observations are purple points, with upper limits denoted by arrows \citep{briley90,lind09}.}
   \label{fig:fl}
\end{figure}

The effect of changing $f_v$ and $f_l$ independently on the surface lithium and carbon abundances is shown in Fig.~\ref{fig:fv} (for $f_v$) and Fig.~\ref{fig:fl} (for $f_l$).

Fast mixing and long mixing lengths (high $f_v$ and $f_l$ respectively) result in increased depletion of both carbon and lithium on the surface. Independently changing $f_v$ and $f_l$ cannot simultaneously match the lithium and the (upper envelope of the) carbon abundances to observations for a single value of $C_t$.

For the case of changing $f_v$ and $f_l$ independently when $C_t = 1000$ the models that best match lithium abundances to observations are $V0.50$ or $L0.33$ (corresponding to ``effective'' $C_t$ values of 500 and 333 respectively). The models that best match carbon abundances to observations are $V1.00$ or $L1.00$ (corresponding to an ``effective'' $C_t$ value of 1000). These results are consistent with the motivation for this study (in \S\ref{subsec:standardmodels}) where if $C_t$ is chosen to match carbon to observations we deplete lithium too much even when uncertainties/errors of the observations are taken into account.

\begin{table*}
\begin{tabular}{ | p{1.5cm}|p{1.5cm}|p{1.5cm}| p{1.5cm}| }
\hline
 Case & $f_v$ & $f_l$ & Name\\
\hline
\multirow{6}{*}{$V$} & 0.10  & 1 & $V0.10$\\
                     & 0.33  & 1 & $V0.33$\\
                     & 0.50  & 1 & $V0.50$\\
                     & 2.00  & 1 & $V2.00$\\
                     & 3.00  & 1 & $V3.00$\\
                     & 10.0  & 1 & $V10.0$\\
\hline
\multirow{6}{*}{$L$} & 1 & 0.10 & $L0.10$\\
                     & 1 & 0.33 & $L0.33$\\
                     & 1 & 0.50 & $L0.50$\\
                     & 1 & 2.00 & $L2.00$\\
                     & 1 & 3.00 & $L3.00$\\
                     & 1 & 10.0 & $L10.0$\\
\hline
\end{tabular}
\caption[Table caption text]{Models tested in Case 1: Independently changing $v$ and $l$ in the two stream advective mixing scheme.}
\label{table:exp1}
\end{table*}

\subsection{Case 2: Changing $v$ and $l$ to maintain constant $D_t$ in the two stream advective mixing scheme}
\label{subsec:exp2results}

Table~\ref{table:exp2} summarises the models tested for Case 2. Here both $f_v$ and $f_l$ are changed to maintain the selected value of $D_t$. We refer to the case letter and then the value of $f_v$ (remembering that $f_l = 1/f_v$). For example, $DV0.10$ refers to the test where $f_v = 0.10$ and $f_l = 10$.

\subsubsection{Results}
\label{subsubsec:multiparm}

\begin{figure}
   \centering
      \includegraphics[width=\columnwidth]{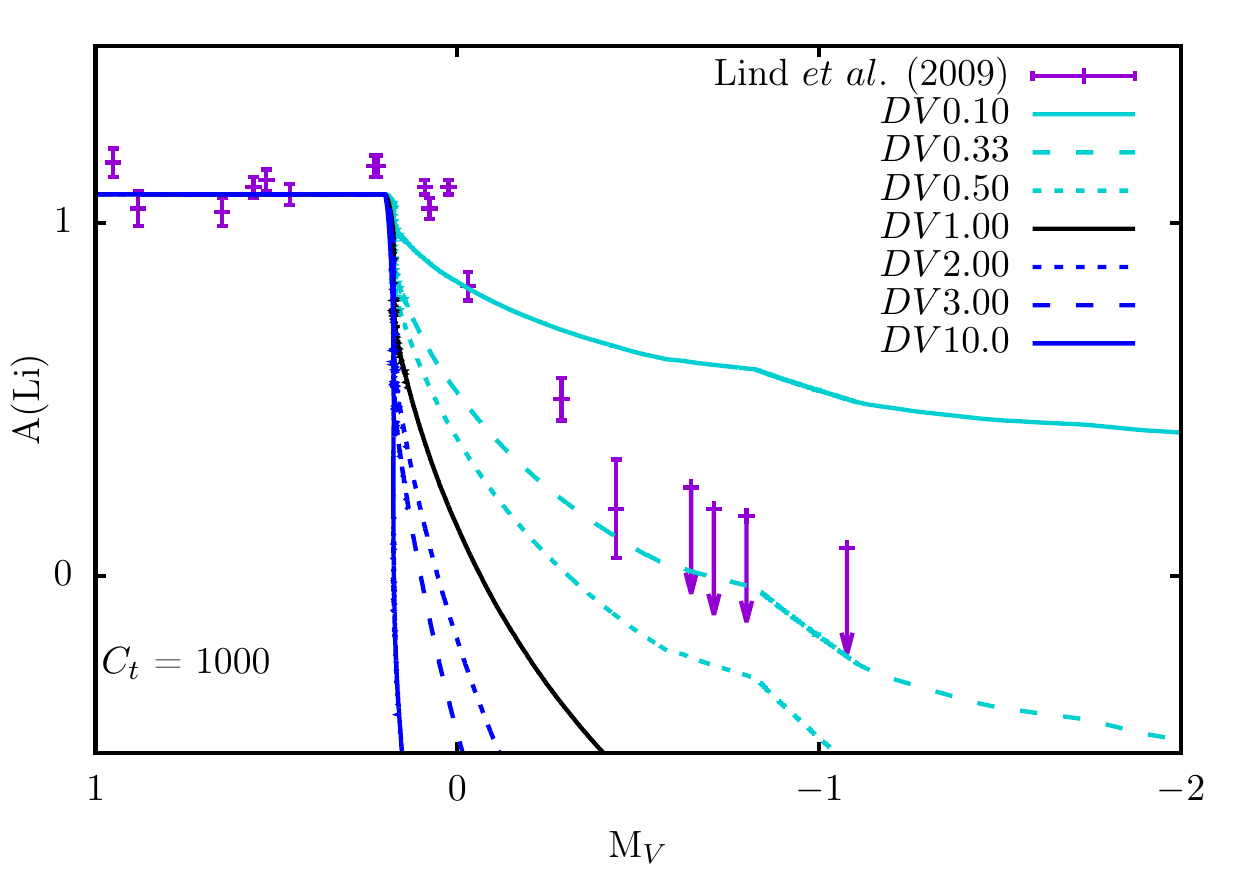}
      \includegraphics[width=\columnwidth]{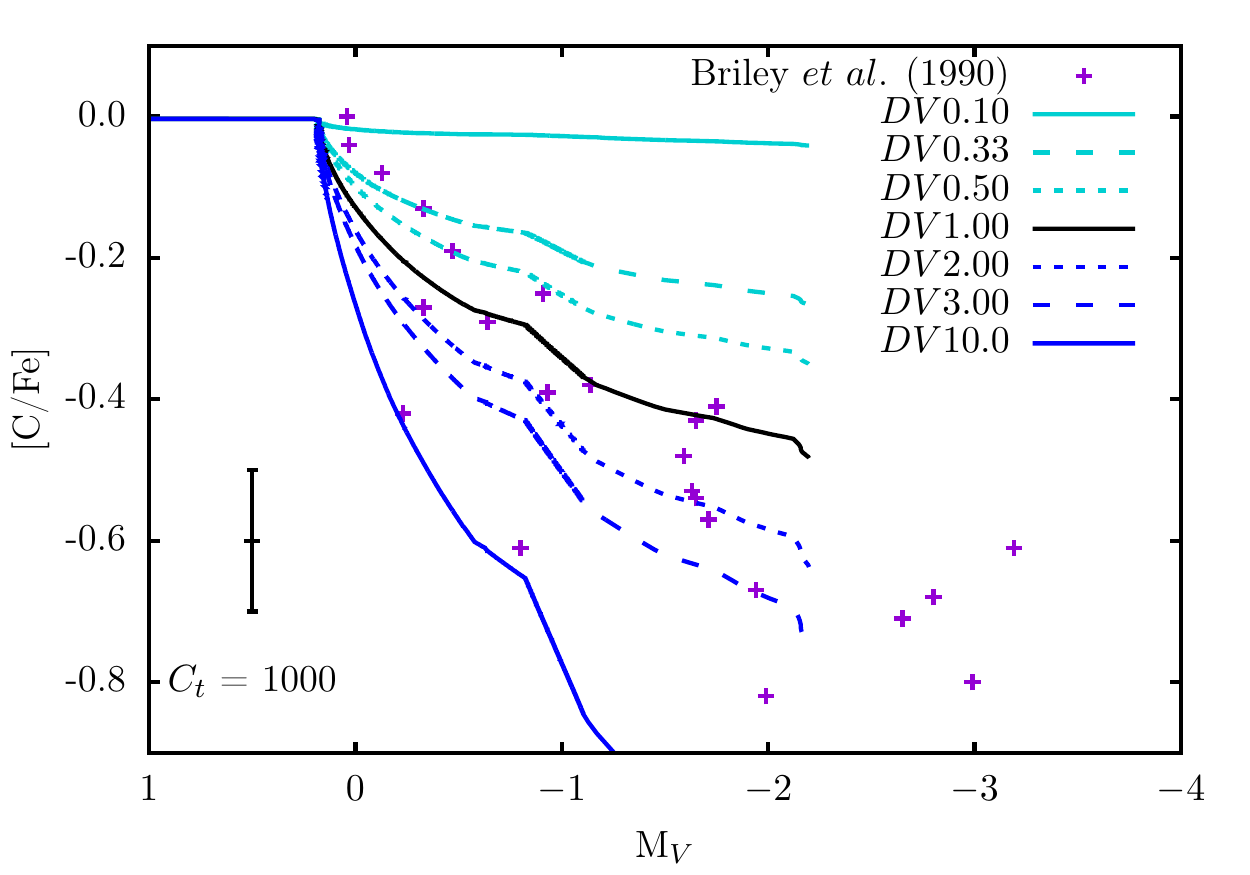}
   \caption{The effect of changing $f_l$ and $f_v$ to maintain constant $D_t$ on the surface abundance profiles of A(Li) (top panel) and [C/Fe] with a 0.1 dex error bar (bottom panel) with $C_t = 1000$. The values of $f_l$ and $f_v$ and colour key of the curves are in the legend of the plots. Observations are purple points, with upper limits denoted by arrows \citep{briley90,lind09}.}
   \label{fig:D}
\end{figure}

Fig.~\ref{fig:D} shows the effect on the surface abundances of lithium (top panel) and carbon (bottom panel). The effect of changing $f_v$ is much more significant than changing $f_l$. To elaborate, varying the velocity of the material has a greater effect on abundances than does varying the mixing length and this is why we see a similar result to changing $f_v$ independently (as in Fig.~\ref{fig:fv}).

The model that best matches models to observations of lithium is $DV0.33$ and the model that best matches carbon is $DV1.00$. This is consistent with the results of independently changing $f_v$ and $f_l$ discussed in \S\ref{subsubsec:singleparm}. In the single parameter tests it was found that models matched observations when $f_v$,$f_l \sim 0.33 - 0.5$ for lithium and when $f_v$,$f_l$ $\sim 1 - 2$ for carbon. We find the same result here. It is evident no solution can be found by modifying $f_v$ and $f_l$ to maintain $D_t$ using our two stream mixing algorithm.

\begin{table*}
\begin{tabular}{ |c | p{1.5cm}|p{1.5cm}| p{1.5cm} |}
\hline
 Case & $f_v$ & $f_l$ & Name\\
\hline
 \multirow{6}{*}{$DV$}  & 0.10 & 10.0 & $DV0.10$\\
                        & 0.33 & 3.00 & $DV0.33$\\
                        & 0.50 & 2.00 & $DV0.50$\\
                        & 2.00 & 0.50 & $DV2.00$\\
                        & 3.00 & 0.33 & $DV3.00$\\
                        & 10.0 & 0.10 & $DV10.0$\\
\hline
\end{tabular}
\caption[Table caption text]{Models tested in Case 2: Changing $v$ and $l$ to maintain constant $D_t$ in the two stream advective mixing scheme.}
\label{table:exp2}
\end{table*}

\subsection{Case 3: Changing $f_u$ and $f_d$ in the two stream advective mixing scheme}
\label{subsec:exp3results}

An important point to consider is that the cross-sectional areas of the streams are unknown; they can and may indeed be unequal in real stars (as opposed to $f_u = f_d = 0.5$). The $f_u$ and $f_d$ values we tested are given in Table~\ref{table:exp3}.

\subsubsection{Results}
\label{subsubsec:exp3}

\begin{figure}
   \centering
      \includegraphics[width=\columnwidth]{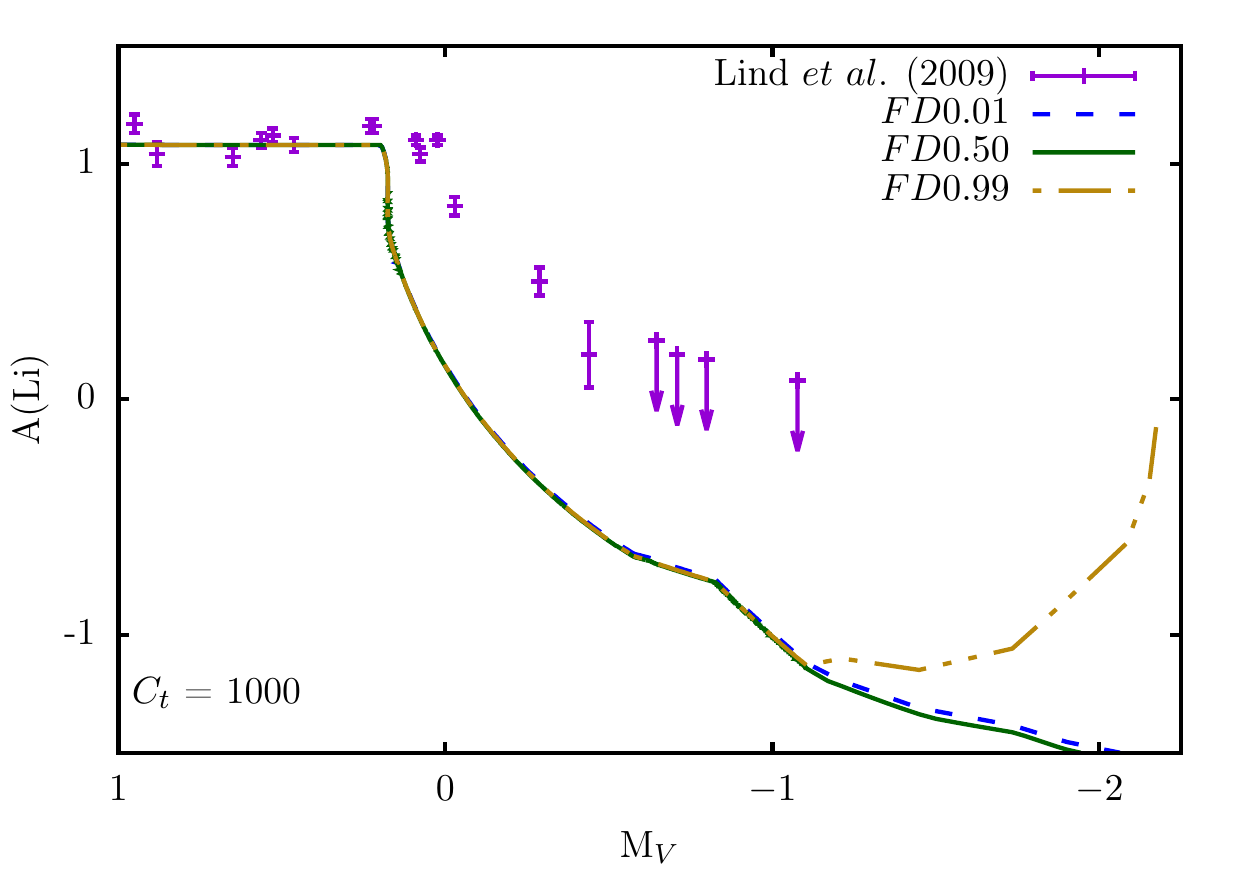}
      \includegraphics[width=\columnwidth]{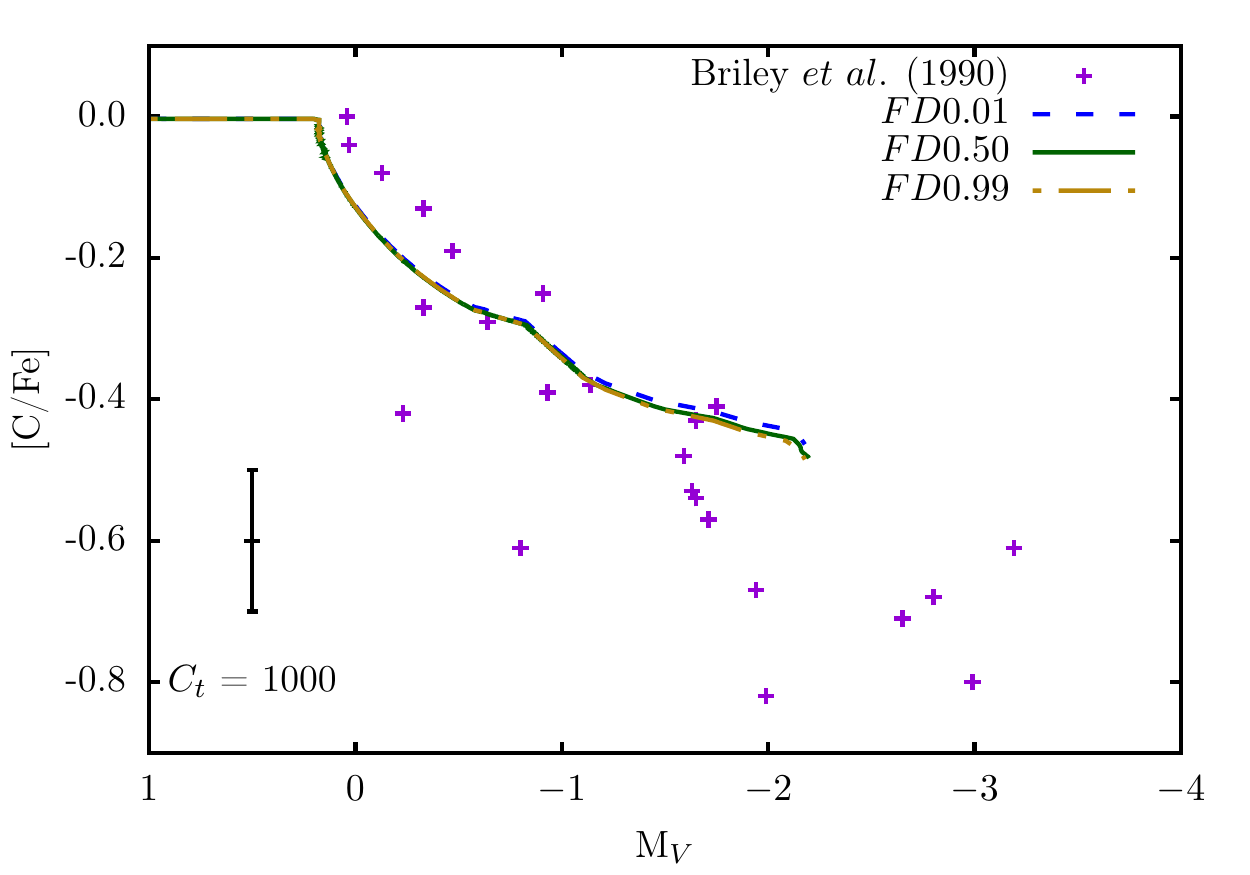}
   \caption{The effect of changing $f_d$ on the surface abundance profiles of A(Li) (top panel) and [C/Fe] with a 0.1 dex error bar (bottom panel) and $C_t = 1000$. The values of $f_d$ and colour key of the curves are in the legend of the plots. Observations are purple points, with upper limits denoted by arrows \citep{briley90,lind09}.}
   \label{fig:fd}
\end{figure}

Fig.~\ref{fig:fd} shows that modifying the up and down stream cross-sectional areas does not significantly affect the results. Indeed, only for the test $FD0.99$ do we see an effect on the lithium abundance, and only towards the end of RGB evolution.

We do no further tests of modifying $f_v$ and $f_l$ (either independently or to maintain $D_t$) for varying $f_d$ because Fig.~\ref{fig:fd} shows that modifying $f_d$ does not have a significant effect on the results and we expect the same result as seen in Figs.~\ref{fig:fv},~\ref{fig:fl}, and~\ref{fig:D} regardless of the value of $f_d$.

\begin{table*}
\begin{tabular}{ |c | p{1.5cm}|p{1.5cm}|p{1.5cm}| }
\hline
 Case & $f_d$ & $f_v$,$f_l$ & Name \\
\hline
\multirow{3}{*}{$FD$} & 0.01 & 1,1 & $FD0.01$ \\
                      & 0.50 & 1,1 & $FD0.50$ \\
                      & 0.99 & 1,1 & $FD0.99$ \\
\hline
\hline
\end{tabular}
\caption[Table caption text]{Models tested in Case 3: Changing $f_u$ and $f_d$ in the two stream advective mixing scheme.}
\label{table:exp3}
\end{table*}

\subsection{Case 4: Changing the thermohaline coefficient in the diffusive mixing scheme}
\label{subsec:exp4results}

The results from \S\ref{subsec:exp1results}, \S\ref{subsec:exp2results} and \S\ref{subsec:exp3results} show that changing the mixing length/velocity in the advective scheme cannot match our models to observations. In the following we develop a method that can ``target'' (in a sense) different elements depending upon their destruction/production timescales.

Recalling that $D_t$ is related to $v$ and $l$ according to Equation~\ref{eq:diff}, when we modify $D_t$ we effectively modify the mixing velocity and/or mixing length. Through the dimensionless value of $C_t$ we can produce different values of $D_t$ to match the observations, which could reveal characteristics of the extra mixing mechanism at the base of the convective envelope during RGB evolution. Manipulating $D_t$ as shown in \S\ref{subsubsec:tempdependence} below expands the parameter space and increases the likelihood of being able to simultaneously match surface carbon and lithium abundances to observations. This may provide us with information on what is missing in the standard implementation. We perform these tests using MONSTAR with mixing modelled by a diffusion equation, as usual (described in \S\ref{subsubsec:thmev}).

\subsubsection{Adding an additional temperature dependence}
\label{subsubsec:tempdependence}

We can manipulate $D_t$ to modify abundances of $^7$Be, $^7$Li, $^{12}$C, and $^{13}$C in the thermohaline region by adding a new temperature dependence on $D_t$. We do this by setting a ``critical temperature'' $T_{\rm crit}$ with ``inner'' $i$ and ``outer'' $o$ factors such that
\begin{equation}
\label{eq:Dnew}
D_{\rm new} = 
   \begin{cases}
      i \times D_t &\text{if } T > T_{\rm crit}, \\
      o \times D_t &\text{if } T < T_{\rm crit}.
   \end{cases}
\end{equation}

From the base of the thermohaline region to the radial location of $T_{\rm crit}$ we multiply the diffusion coefficient $D_t$ as given in Equation~\ref{eq:Dthm} by a factor $i$. Similarly, from the location of $T_{\rm crit}$ to the envelope we multiply $D_t$ by a factor $o$.

The value of $T_{\rm crit}$ dictates the elements that are effected by $i$ and $o$. To elaborate, if a ``high'' $T_{\rm crit}$ value is selected so that it is sufficiently close to the base of the thermohaline region then $i$ will only affect elements burning at the highest temperatures and $o$ will have a larger effect on $D_t$ (and all abundances regardless of burning temperature). If $T_{\rm crit}$ is ``low'' and located sufficiently close to the envelope then $o$ will have little effect. If $T_{\rm crit}$ is somewhere in the middle of the thermohaline region then $i$ will predominantly affect the abundances of elements that have high burning temperatures ($^{12}$C and $^{13}$C) and $o$ will predominantly affect the abundances of elements with low burning temperatures ($^7$Be and $^7$Li).

\subsubsection{Results}
\label{subsubsec:exp4}

\begin{figure}
   \centering
      \includegraphics[width=\columnwidth]{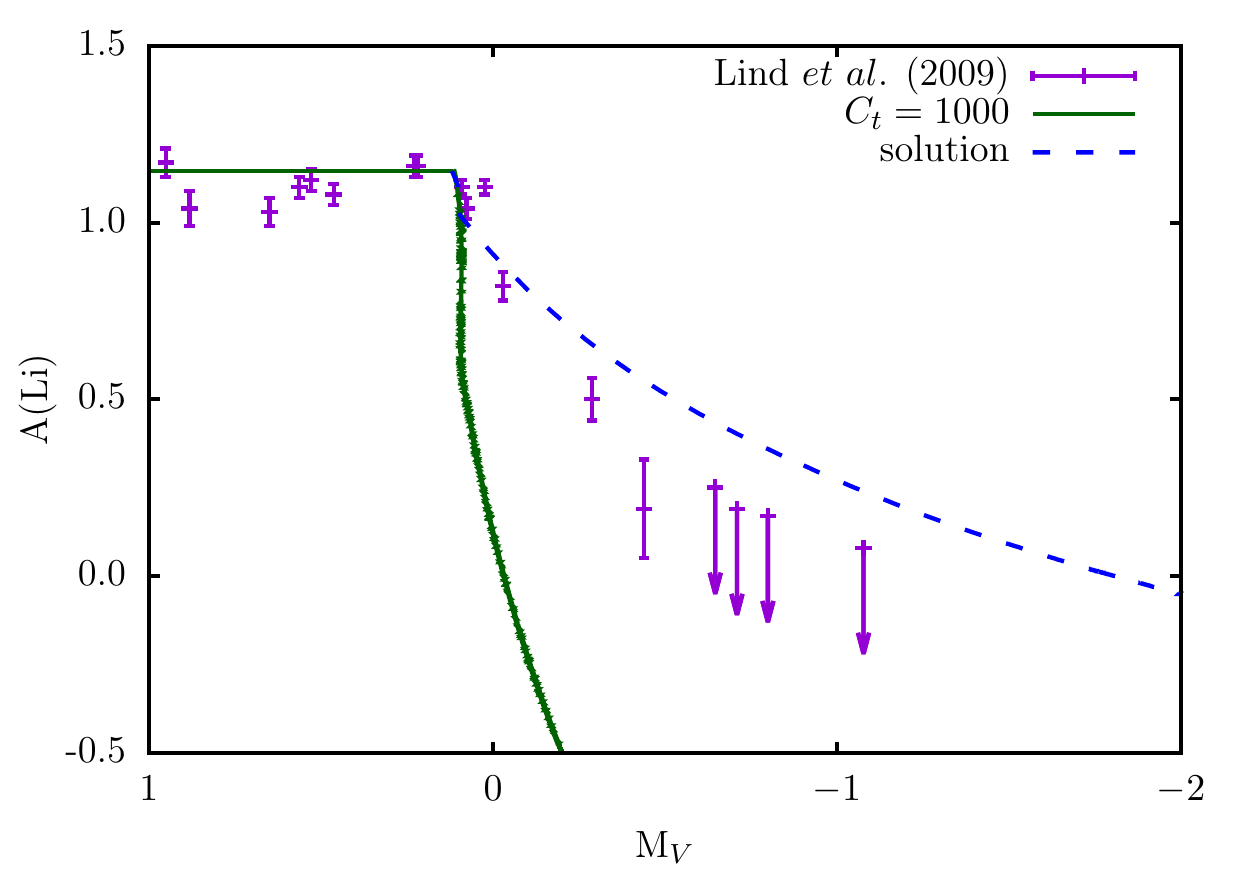}
      \includegraphics[width=\columnwidth]{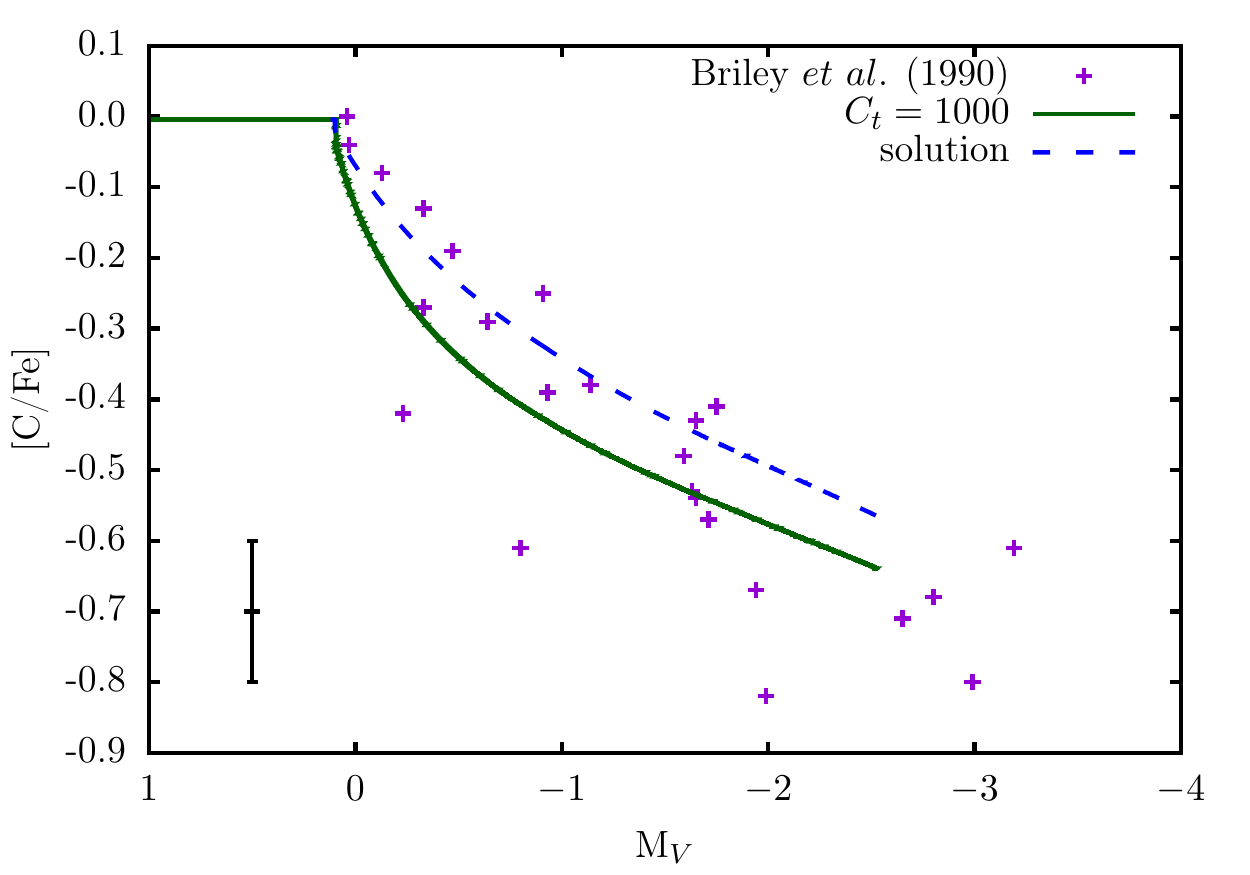}
   \caption{Top panel: Theoretical surface lithium abundances (curves) compared to observations with errors \citep[purple points, arrows indicate upper limits,][]{lind09}. Bottom panel: Theoretical carbon abundances (curves) compared to observations with a 0.1 dex error bar \citep[purple points,][]{briley90}. In each panel the dark green curve is when the thermohaline mixing free parameter $C_t = 1000$. The blue dashed curve is one solution with $T_{\rm crit} = 8$MK, $i = 3.0$, and $o = 0.1$.}
   \label{fig:8mki3o01errors}
\end{figure}

\begin{figure}
   \centering
      \includegraphics[width=\columnwidth]{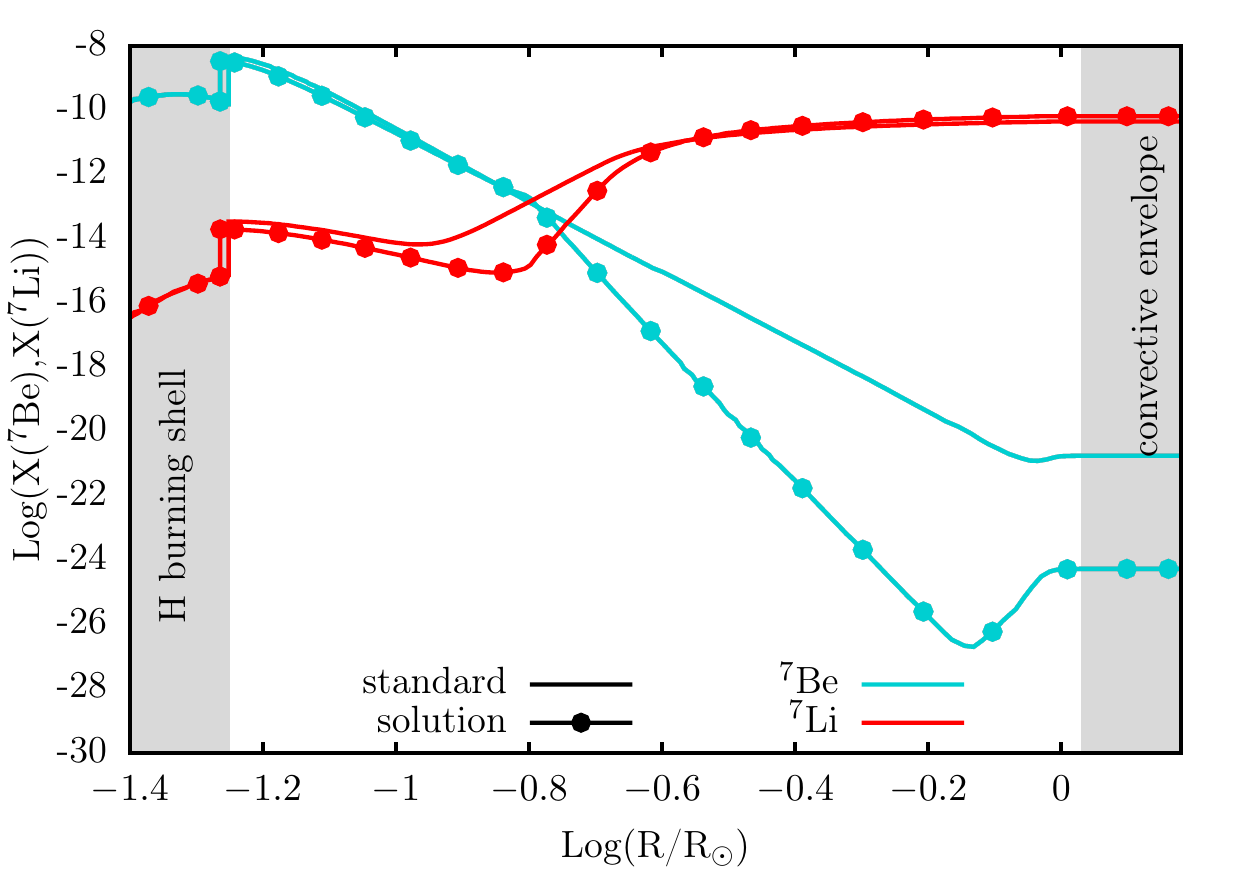}
      \includegraphics[width=\columnwidth]{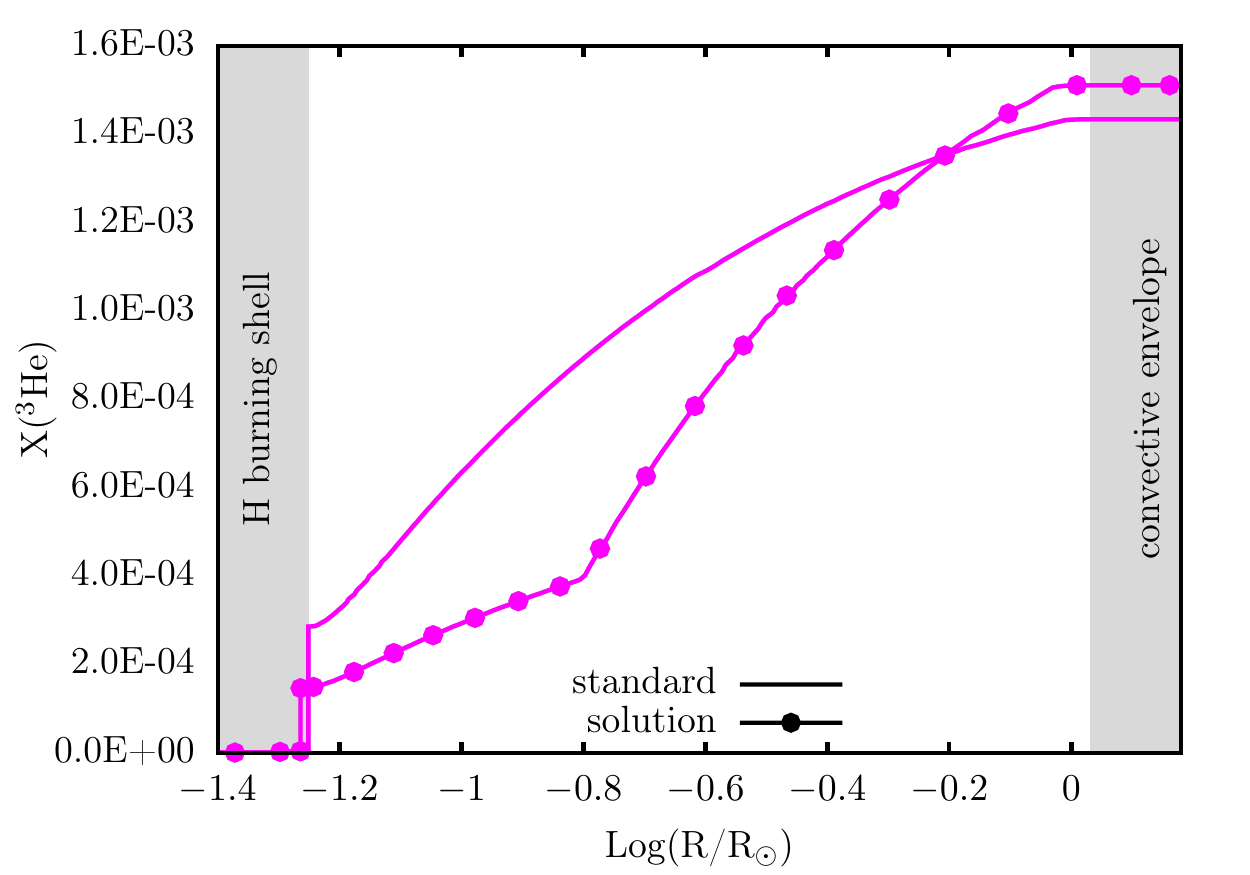}
      \includegraphics[width=\columnwidth]{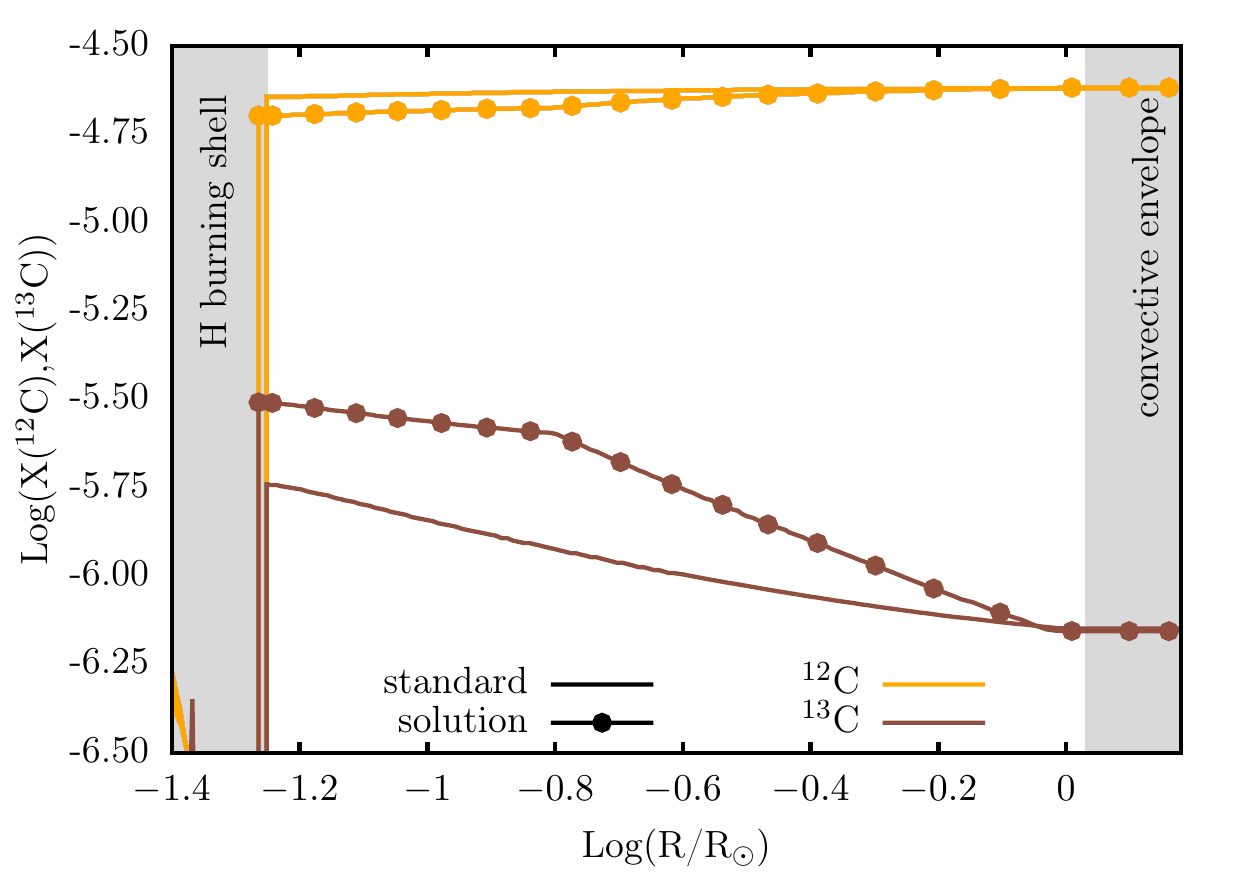}
   \caption{Abundance profiles for $f$ = 1 standard case (solid curves), and one solution where $T_{\rm crit} = 8$MK, $i = 3.0$ and $o = 0.1$ (filled circles). Top panel: $^7$Li (red) and $^7$Be (turquoise). Bottom panel: $^{12}$C (orange) and $^{13}$C (burgundy).}
   \label{fig:8mki3o01elements}
\end{figure}

The number of combinations of $D_t$, $T_{\rm crit}$, $i$, and $o$ available mean that there are a family of solutions that can simultaneously match carbon and lithium abundances to observations. An analytic multi-dimensional theory of thermohaline mixing is not available and beyond the scope of this paper. Hence in the first instance we seek solutions that are within one order of magnitude of the diffusion coefficient as found by the Ulrich/Kippenhahn 1D theory. In Figs.~\ref{fig:8mki3o01errors} and~\ref{fig:8mki3o01elements} we show one solution where $T_{\rm crit} = 8 \text{MK}$, $i = 3.0$ and $o = 0.1$. We could certainly generate a better fit, but since the modification we used is purely phenomenological we feel that this would not add any insights. The present result tells us that to match both carbon and lithium simultaneously we need faster mixing in hot regions and slower mixing in cooler regions.

\section{Discussion and Conclusions}
\label{sec:discussion}

We have modified the thermohaline mixing model to be able to match lithium and carbon abundances to observations of NGC 6397 red giants. We do this by adding an additional temperature dependence to the thermohaline diffusion coefficient. Our successful model proposes mixing that is faster than the standard 1D theory predicts near the base of the thermohaline region, and mixing that is slower further out in the region towards the base of the convective envelope.

The effect of our modification to the standard theory is to facilitate the Cameron-Fowler mechanism by reducing the decline of the lithium surface abundance. Our model achieves this by removing beryllium faster from the vicinity of the H shell and allowing it to remain longer in the outer part of the radiative zone (where it captures an electron to produce lithium). A similar effect can be obtained when one uses a constant diffusion coefficient, which was shown by \citet{denissenkov03}.

Our successful method results in a number of non-unique solutions because of the size of the parameter space available. Despite this success, there are caveats that must be noted. Our model is 1D and subject to the uncertainties and limitations that are inherent in all stellar evolution modelling \citep[e.g.,][]{karakas14dawes}. Also, we are adding an additional temperature dependence that is not yet driven by physics. There may be physics in the 1D theory derived by \citet{ulrich72} and \citet{kippenhahn80} that could drive such an extra temperature dependence that has not been identified. This is an avenue for further research and beyond the scope of this paper.

Modifying our advective scheme mixing length and velocity parameters independently (producing a change in $D_t$) and dependently (maintaining constant $D_t$) could not simultaneously match carbon and lithium for the same set of parameters. Our advective regime, unlike a diffusion equation, does not mix along a composition gradient and all elements present in the mixing regions are carried in the streams with the same velocity as the streams themselves. Fast mixing and long mixing lengths (high $f_v$ and $f_l$ respectively) result in increased depletion of both carbon and lithium on the surface because they, and all other elements in the mixing region, are brought down more quickly to high-temperature regions where they are burnt. Changing the velocity of the streams and/or the mixing length does not affect certain elements differently to others, therefore carbon and lithium are depleted more when the velocity and mixing length are increased (and vice versa when the mixing velocity and length are decreased). This is the limitation of this method and is the main reason why it is unsuccessful. Indeed, the standard diffusion implementation also fails because it shows these characteristics.

We look to results from multi-dimensional studies to inform 1D stellar modelling. The theory derived by \citet{ulrich72} is one-dimensional. Further, it cannot adequately constrain the aspect ratio ($\alpha$ in Equation~\ref{eq:Cthm}) of the fingers. Studies using 3D simulations of surface \citep{robinson03,steffen05,collet07} and interior \citep{stancliffe11,ohlmann16} convection zones in stars have found that the upstream velocity is slower and the cross-sectional area of the upstream flow is larger than the respective values for the downstream, i.e. $v_u < v_d$ and $f_u > f_d$. These effects do not appear in the standard theory.

Multi-dimensional simulations of thermohaline mixing have also been performed. The 2D simulations of \citet{denissenkov10} show that fingers (corresponding to $\alpha \sim 7$) of material arise in the oceanic thermohaline environment but blobs with $\alpha \sim 0.5$ occur in the RGB case. \citet{denissenkov10} achieved fingers of material (with $\alpha > 1$) in their RGB case only for highly viscous environments (viscosities that are 4 orders of magnitude higher than in real RGB stars). \citet{garaud15} found shearing in their 2D simulations when the Prandtl number was less than 0.5 that was not seen in their 3D simulations, and concluded that for sufficiently low Prandtl numbers, 3D models are necessary to resolve the thermohaline environment.

Several other 3D simulations of thermohaline mixing have been conducted in recent years \citep{denissenkov11,traxler11,brown13,medrano14,garaud15}. \citet{denissenkov11} compared their 3D work to the 2D simulations of \citet{denissenkov10} and confirmed the results from the 2D simulations, as well as finding the excitation of gravity waves in their oceanic case but not their RGB case. Secondary instabilities (e.g., gravity waves) triggered by thermohaline mixing have been found by other groups \citep{traxler11,garaud15}. Three-dimensional simulations show that the shape of the thermohaline fingers changes as conditions come closer to representing real stellar conditions (i.e. as the Prandtl number decreases), with the thermohaline fingers becoming more like blobs \citep{traxler11}. Subsequent studies found that thermohaline fingers became blobs over time \citep{brown13,medrano14,garaud15}.

In all of the multi-dimensional thermohaline mixing simulations above, the sizes of the down and upstreams were approximately equal to each other, and we found from our 1D study no significant effect when the stream size was altered. However, the simulated stellar environments are not entirely representative of the conditions in real stars. Real stars are likely to be much more turbulent because the Prandtl number is extremely small in reality ($\sim 10^{-6}$) compared to simulations, which typically have Prandtl numbers $\sim 0.1 - 0.01$ \citep{traxler11,brown13,garaud15}. Additionally, the density ratio, the ratio of the (stabilising) entropy gradient to the (destabilising) compositional gradient, in simulations is generally $\sim 1.1$ and much lower than the RGB value of $1.7 \times 10^3$ \citep{traxler11,brown13,garaud15}. \citet{brown13} found that a density ratio of 1.1 resulted in larger, convective-like plumes, whereas a more turbulent environment with a density ratio of 3 was more finger-like, indicating that the density ratio in real RGB stars could produce fingers of material as opposed to blobs.

It is clear that we do not yet adequately understand the thermohaline mechanism. Additionally, to provide theoretical stellar models sufficient (observational) constraints, more observations of both carbon and lithium in globular cluster red giants are needed for stars in the same cluster (other than NGC 6397) covering a range in metallicity. This will help us to determine what the implementation we used to match models to NGC 6397 is telling us about necessary modifications to the standard theory.

\section*{Acknowledgements}

K.H. acknowledges the financial support of the Australian Postgraduate Award scholarship and would like to thank the Stellar Interiors and Nucleosynthesis (SINs) group at Monash for their helpful discussions.

The authors thank the referee for their useful suggestions.



\bibliographystyle{apj}
\bibliography{mnemonic,library}






\bsp
\label{lastpage}
\end{document}